\documentclass[sigconf,screen]{acmart}

\AtBeginDocument{%
  }

\copyrightyear{2025}
\acmConference[Author's Version]{}{}{}

\usepackage{lipsum}
\usepackage{wrapfig}
\usepackage{graphicx}
\usepackage{subcaption}
\usepackage{caption}
\usepackage{enumitem}
\usepackage{pdflscape}
\usepackage{url}

\setitemize{noitemsep,topsep=0pt,parsep=0pt,partopsep=0pt}
\newcommand{\revision}[1]{#1}

\newcommand{\topic}[1]{\textbf{#1}}

\newcommand{\Constant}{\textsc{Constant}}
\newcommand{\GoGo}{\textsc{Go-Go}}
\newcommand{\PRISM}{\textsc{PRISM}}
\newcommand{\ForcePinch}{\textsc{ForcePinch}}

\begin{document}

\title[ForcePinch: Force-Responsive Spatial Interaction]{ForcePinch: Force-Responsive Spatial Interaction for \\ Tracking Speed Control in XR}

\author{Chenyang Zhang}
\email{chenyang.zhang@gatech.edu}
\orcid{0009-0003-1116-4895}
\affiliation{
  \institution{Georgia Institute of Technology}
  \city{Atlanta}
  \state{Georgia}
  \country{USA}
}

\author{Tiffany Ma}
\email{tma304@gatech.edu}
\orcid{0009-0006-2119-0183}
\affiliation{
  \institution{Georgia Institute of Technology}
  \city{Atlanta}
  \state{Georgia}
  \country{USA}
}

\author{John Andrews}
\email{jandrews304@gatech.edu}
\orcid{0009-0002-8898-7681}
\affiliation{
  \institution{Georgia Institute of Technology}
  \city{Atlanta}
  \state{Georgia}
  \country{USA}
}

\author{Eric Gonzalez}
\email{ejgonz@google.com}
\orcid{0000-0002-2846-7687}
\affiliation{
  \institution{Google}
  \city{Seattle}
  \state{Washington}
  \country{USA}
}

\author{Mar Gonzalez-Franco}
\email{margon@google.com}
\orcid{0000-0001-6165-4495}
\affiliation{
  \institution{Google}
  \city{Seattle}
  \state{Washington}
  \country{USA}
}

\author{Yalong Yang}
\email{yalong.yang@gatech.edu}
\orcid{0000-0001-9414-9911}
\affiliation{
  \institution{Georgia Institute of Technology}
  \city{Atlanta}
  \state{Georgia}
  \country{USA}
}

\renewcommand{\shortauthors}{Zhang et al.}

\begin{abstract}
Spatial interaction in 3D environments requires balancing efficiency and precision, which requires dynamic tracking speed adjustments. However, existing techniques often couple tracking speed adjustments directly with hand movements, reducing interaction flexibility. Inspired by the natural friction control inherent in the physical world, we introduce \textbf{ForcePinch}, a novel force-responsive spatial interaction method that enables users to intuitively modulate pointer tracking speed and smoothly transition between rapid and precise movements by varying their pinching force. To implement this concept, we developed a hardware prototype integrating a pressure sensor with a customizable mapping function that translates pinching force into tracking speed adjustments. We conducted a user study with 20 participants performing well-established 1D, 2D, and 3D object manipulation tasks, comparing ForcePinch against the distance-responsive technique Go-Go and speed-responsive technique PRISM. Results highlight distinctive characteristics of the force-responsive approach across different interaction contexts. \revision{Drawing on these findings, we highlight the contextual meaning and versatility of force-responsive interactions through four illustrative examples, aiming to inform and inspire future spatial interaction design.}
\end{abstract}



\begin{teaserfigure}
    \vspace{-1.5em}
    \centering
    \begin{subfigure}{0.45\textwidth}
        \includegraphics[width=\linewidth]{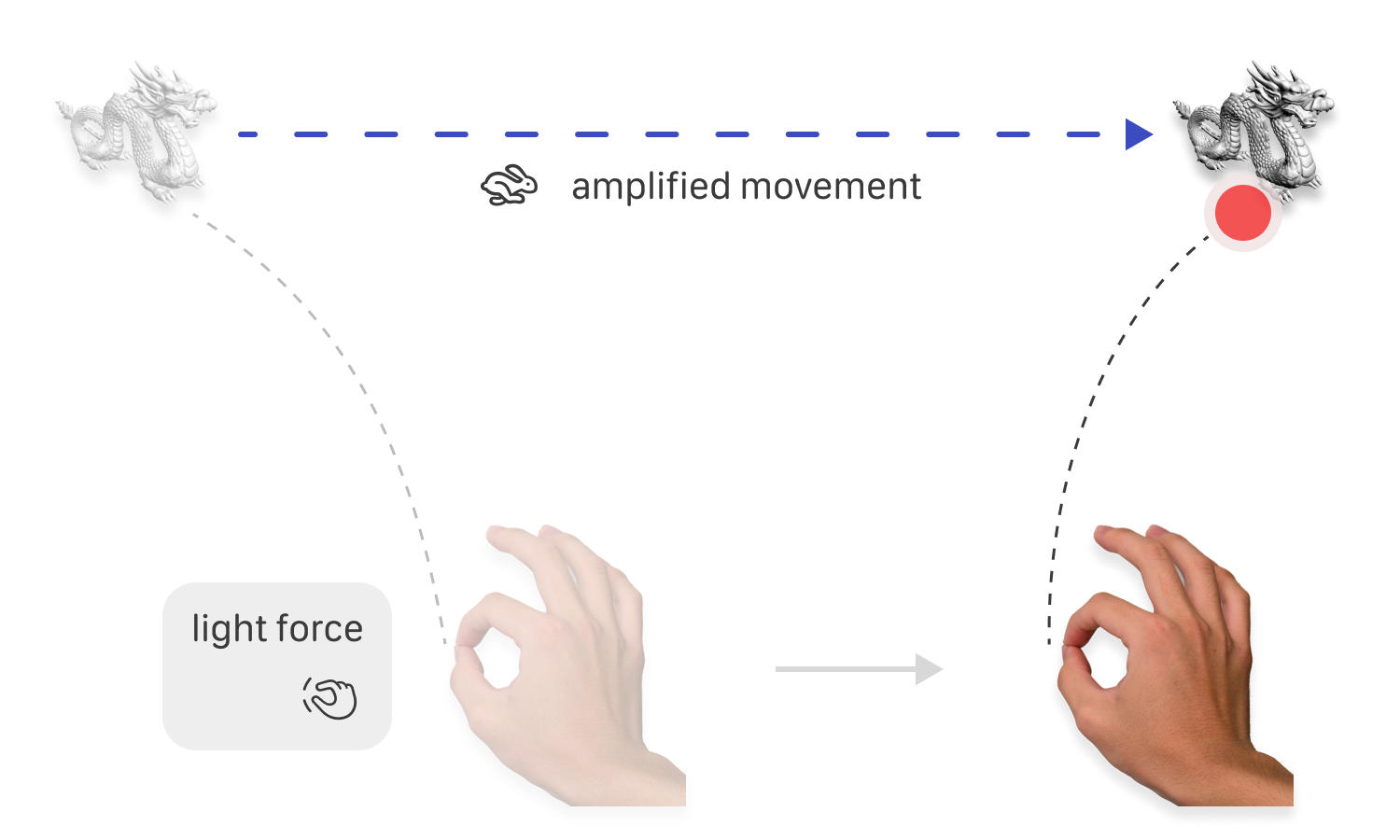}
        \caption{Fast Tracking Speed}
        \label{fig:teaser-fast}
    \end{subfigure}
    \hspace{1.0cm}
    \begin{subfigure}{0.45\textwidth}
        \includegraphics[width=\linewidth]{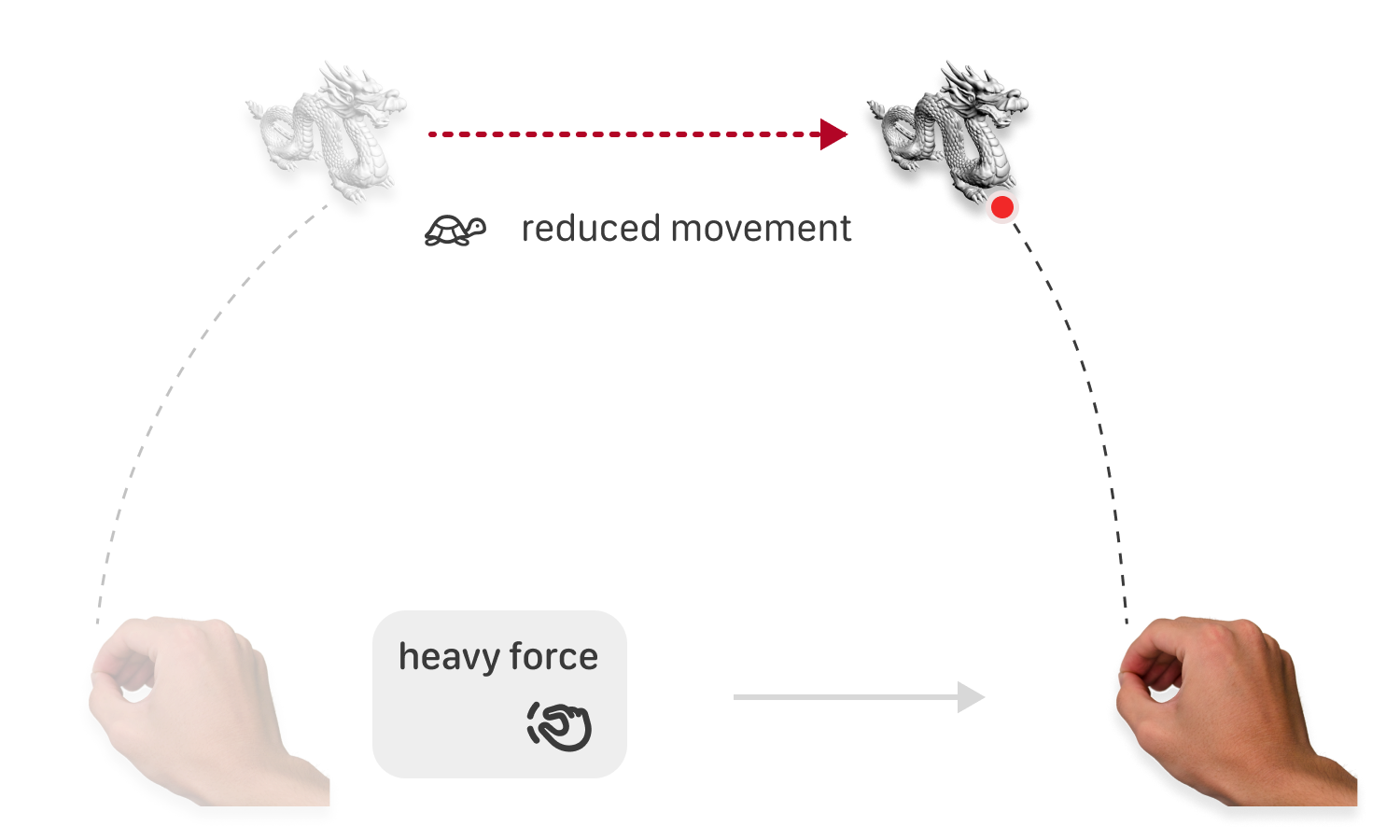}
        \caption{Slow Tracking Speed}
        \label{fig:teaser-slow}
    \end{subfigure}
    \vspace{1.2em}
    \caption{ForcePinch is an object manipulation method that allows users to control the tracking speed of a mid-air pointer \raisebox{-0.2em}{\includegraphics[height=0.9em]{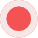}} by modulating pinch force during object manipulation. (a) A light pinch results in fast tracking, supporting broad and rapid movements. (b) A heavy pinch produces slow tracking, allowing for fine-grained precision.}
    \vspace{1.2em}
    \label{fig:teaser}
    \Description{This figure shows how ForcePinch maps pinch force to tracking speed. In (a), a light pinch enables fast tracking for broad, rapid movement. In (b), a heavy pinch slows the tracking, allowing for precise adjustments.}
\end{teaserfigure}

\maketitle

\begin{center}
\fbox{\parbox{0.95\linewidth}{%
\textbf{Author’s Version.} This is the author's version of the work. The definitive version is published in the \textit{Proceedings of the 38th Annual ACM Symposium on User Interface Software and Technology (UIST '25)}. DOI: \url{https://doi.org/10.1145/3746059.3747694}
}}
\end{center}

\section{Introduction}

Immersive technology allows us to seamlessly integrate digital objects into our physical surroundings, enabling users to interact and manipulate virtual content directly within the 3D space. However, when interacting with digital objects, users often encounter conflicting control schemes depending on the specific scenario. Sometimes, users prioritize rapid, large-scale interactions, such as swiftly moving a virtual book from a nearby table to a distant bookshelf. In other scenarios, precision becomes essential, such as meticulously aligning a small virtual object to some desired position. Their expectations of the control scheme may differ from the actions to execute.

Fundamentally, when interacting with objects in a 3D virtual environment, there is a critical trade-off regarding how physical hand movements are translated into virtual object movements. This relationship is typically described by the concept of \emph{tracking speed} (or control-display ratio), defined as how quickly or slowly virtual pointer movements respond to the user's physical hand movements. Higher tracking speeds allow users to cover large virtual distances quickly with small hand movements, enhancing interaction efficiency but typically compromising precision. Conversely, lower tracking speeds facilitate precise and controlled object manipulation, but reduce users' ability to move virtual objects rapidly across larger spaces. Many interaction tasks require dynamically adjusting the tracking speed between these extremes. For instance, users might initially benefit from a higher tracking speed when quickly grabbing a distant object, then prefer a lower tracking speed to accurately position it at a desired location. Thus, there is a compelling need for intuitive methods that allow users to flexibly and dynamically adjust the tracking speed according to their immediate interaction needs.


To address the balance between interaction speed and precision, several adaptive tracking speed control methods have been proposed for spatial interactions.
State-of-the-art techniques like Go-Go \cite{poupyrev1996go} and PRISM \cite{frees2007prism} adjust pointer tracking speed based on hand movement characteristics—specifically, Go-Go increases reach based on distance from the body, while PRISM adjusts sensitivity according to hand speed.
However, these methods inherently couple sensitivity control with hand movement characteristics, potentially causing inaccuracies like overshooting \cite{casiez2008impact}. Other techniques such as dual-precision \cite{nancel2013high} and heuristic pointing \cite{de2005intenselect} adapt tracking speed based on context, but users' control is limited, and their precision in complex scenarios remains questionable.


In this work, we introduce \textbf{ForcePinch}, a force-responsive interaction method where pointer tracking speed directly responds to the user's pinch force. ForcePinch enables users to dynamically control tracking speed independently of hand movement, thus facilitating natural and efficient object manipulation in mixed reality. To establish this interaction, we first constructed a hardware prototype incorporating a pressure sensor to precisely measure the pinching force \(f\) between thumb and index finger. Next, we defined a negatively correlated mapping function between the pinch force \(f\) and pointer tracking speed \(s\). This mapping function is customizable to accommodate individual variations in users' strength and sensitivity perceptions. To further enhance interaction awareness, we provide real-time visual feedback reflecting changes in pinch force. We evaluated ForcePinch's performance and usability through a comprehensive user study involving 20 participants across various tasks in 1D, 2D, and 3D scenarios, comparing our approach against baseline methods such as Go-Go and PRISM. Our results indicate that ForcePinch successfully enables tracking speed control and notably increase the precision and user experience in 1D and 2D tasks, but may reduce its effectiveness in 3D placement.


\revision{
Alongside our implementation of ForcePinch, we observed a broader opportunity for integrating continuous pinch force into spatial embodied interactions, such as selection~\cite{lu2020investigating} or text input~\cite{bowman2002text}. While our study focuses on one mapping—modulating tracking speed via pinch force—we reflect on how continuous force input can act as a secondary channel to enrich a variety of primary interactions. We articulate this idea under the term \textbf{force-responsive interaction}, where continuous force modulation augments or controls core interaction tasks like pointing, manipulating, or drawing. Rather than proposing a formal framework, we present this concept as a lens to inspire future design and highlight promising directions for more expressive, adaptive interaction techniques, as discussed in Section~\ref{sec:reflection}.
}

The primary contributions of our work include:
\begin{enumerate}[leftmargin=*]
    \item ForcePinch, a novel spatial interaction technique that leverages pinching force as a modifier, enabling natural and dynamic control of tracking speed.
    \item An evaluation of ForcePinch for object translation, comparing its performance with baseline methods and analyzing its advantages and limitations across different scenarios.
    \item \revision{A reflection on the broader design potential of force-responsive interaction, illustrating how continuous force input can complement primary embodied tasks in XR.}
\end{enumerate}

\section{Related Work}

\subsection{Force Input and Interaction}
In screen-based interactions, for example, systems like BackXPress~\cite{corsten2017back} and PreSenseII~\cite{rekimoto2006presense} utilize force sensitivity on touchscreens or device backs to enable new quasi-modes and multilevel input states. 
Force input has also been coupled with haptic feedback, as seen in deformable interfaces like Haptic Chameleon~\cite{michelitsch2004haptic} and X-Rings~\cite{gonzalez2021xrings}, allowing users to dynamically manipulate shape-changing controls. 
While these applications demonstrate innovative uses of force for mode switching or tangible feedback, their operational contexts differ significantly from mid-air spatial interaction, and thus they do not directly address the specific challenge of fluidly balancing speed and precision for object manipulation in AR/VR.

Turning to wearable interfaces, particularly those focused on finger interactions, force sensing provides a nuanced control channel beyond traditional touch and gesture inputs. 
Early explorations often integrated force sensing into compact form factors. 
MicroPress, for instance, leverages thumb-to-finger pressure for discrete tasks such as pressure-sensitive text entry and selection~\cite{dobinson2022micropress}. Similarly, ElectroRing employs a minimal ring design to detect subtle micro-gestures and pressure variations between fingers~\cite{kienzle2021electroring}. 
Although these works successfully demonstrated the feasibility of finger-based force input, their emphasis was often on gesture recognition or triggering discrete actions. 
Consequently, they less commonly leveraged the continuous range of applied force for dynamic, proportional control over interaction parameters like movement speed, representing a gap that ForcePinch aims to address.

Building upon these foundations within the specific context of VR and AR, force sensing has been further investigated to enhance immersion and control. 
Techniques vary, including inferring force indirectly from muscle activity via EMG~\cite{zhang2024may}, or integrating direct pressure sensing into gloves or rings to support tasks like hand-eye coordination (DigiTouch, NailRing, GazeRing)~\cite{whitmire2017digitouch, li2022nailring, wang2024gazering} or facilitate text editing through finger swipes (Swipe-It~\cite{kim2023swipe}, ThumbText \cite{kim2018thumbtext}). 
\revision{
DeformWear~\cite{weigel2017deformwear} is another notable system—a flexible hardware platform that supports a variety of deformation gestures. While DeformWear offers a general-purpose deformable input surface, our work focuses specifically on pinch gesture-based speed modulation, introducing a distinct interaction primitive.
}
These studies effectively showcase the value of incorporating pressure modulation into immersive interactions for various applications. 
However, they often target these specific use cases or utilize force as just one component within more complex interaction schemes. 
This leaves an opportunity, explored by ForcePinch, to investigate how direct, continuous pinch force itself can serve as an intuitive, primary mechanism for modulating a fundamental spatial interaction parameter like object tracking speed during manipulation.

\subsection{Tracking Speed Control}

Adaptive mapping techniques are crucial for improving precision and efficiency in pointer-based interactions by dynamically modulating the control-display (C-D) gain. 
While common in desktop environments where pointer acceleration adjusts gain based on movement speed, this can risk overshoot if miscalibrated~\cite{Casiez2008gain}. 
In 3D interactions, managing C-D gain is even more critical. The foundational Go-Go technique addressed large interaction volumes by introducing non-linear distance-to-reach mapping, extending reach without explicit modes~\cite{poupyrev1996go}. 
However, being primarily distance-triggered, Go-Go's gain adaptation is automatic based on reach distance, not offering users fine-grained, intentional control over the gain based on immediate task needs. 
Nevertheless, this principle of modeless adaptation based on motion characteristics shaped many subsequent systems.

Building upon this, subsequent techniques refined dynamic gain adjustment, often tying it to other motion parameters. 
PRISM, for instance, introduced velocity-based scaling, reducing gain for slow movements (assumed precise) and increasing it for fast ones~\cite{frees2007prism}, a concept integrated into systems like Scaled HOMER~\cite{Wilkes2008velocity}. 
Others focused on ergonomic transitions, like Reach-Bounded Nonlinear amplification~\cite{Wentzel2020ergonomics}. While effective in improving accuracy and reducing jitter compared to fixed gain, these methods still primarily tie gain modulation to inherent motion characteristics like velocity or reach distance. 
This limits the user's ability to intentionally decouple their physical movement speed from the resulting cursor speed. 
The gain adaptation remains largely an inferred consequence of movement, not a directly controlled parameter.

\revision{
Force-to-Motion~\cite{rutledge1990force} explores pressing force via a joystick-like device to modulate 2D cursor control in desktop settings. In contrast, our work uses two-finger pinch force in embodied 3D space, introducing different design and control challenges related to proprioception, spatial coordination, and mid-air precision. These differences underscore the unique affordances and constraints of force input in spatial interaction contexts.
}

More recent work explores richer input contexts and multimodality. PinchLens cleverly uses a semi-pinch gesture to activate a magnified selection mode and simultaneously reduce C-D gain, aiding precise targeting~\cite{zhu2023pinchlens}. 
Depth-adaptive cursors infer user intent based on context like proximity and viewpoint to adjust the cursor~\cite{zhou2022mouse}. 
While innovative, these approaches often link gain changes to discrete gestures or rely on system inference, rather than providing continuous, proportional control over gain itself. 

Thus, a gap remains in exploring how a continuous, intuitive input channel, separate from primary motion or discrete gestures, can provide direct, real-time, proportional modulation of tracking gain. 
ForcePinch addresses this specific gap by proposing applied pinch force as such a mechanism, allowing users to intuitively adjust control granularity during object manipulation independently of their hand's motions.

\subsection{3D Interaction}

Effective 3D interaction techniques are fundamental to realizing the potential of VR/AR, enabling users to manipulate virtual objects naturally and efficiently \cite{laviola20173d}. 
Given the complexity of interacting within a 3D space, early research focused on establishing foundational understanding and classification. 
Frameworks were developed to categorize interaction techniques based on task requirements and input modalities, providing a structured way to analyze and compare different approaches~\cite{poupyrev1997framework, hertel2021taxonomy}. 
These taxonomies highlighted the importance of considering factors like selection, manipulation, and navigation, often emphasizing the potential benefits of multimodal input for enhancing usability.

Building on these frameworks, research extensively developed and evaluated specific techniques, particularly for remote object manipulation. Comparative studies, like Bowman et al.'s evaluation of methods such as Go-Go and Ray-Casting~\cite{bowman1997evaluation}, revealed critical trade-offs between precision, speed, and ease of use, underscoring that technique choice is context-dependent. 
Furthermore, studies on interaction fidelity highlighted the need to balance desired realism with practical usability to ensure efficiency and avoid fatigue~\cite{rogers2019exploring}.

Consequently, much research has focused on refining interaction techniques to overcome inherent limitations, particularly the challenge of balancing speed and precision across varying distances. 
Strategies like adaptive control-display gain~\cite{liu2022distant}, dynamic speed control~\cite{gemici2024object}, and near-field scaled manipulation~\cite{babu2024benefits} aim to enhance precision without sacrificing rapid movement capabilities. 
Alongside these refinements, alternative input modalities like gaze-supported manipulation~\cite{yu2021gaze} and tangible proxies~\cite{englmeier2020tangible} have been explored to improve selection and control.

Our work contributes to this ongoing effort by investigating force-based input. 
Specifically, we explore continuous pinch force as an intuitive mechanism to modulate interaction parameters like tracking speed, aiming to provide fine-grained control and address the persistent challenge of seamlessly balancing speed and precision during 3D object manipulation in immersive environments.
\begin{figure*}
    \centering
    \begin{subfigure}{0.74\linewidth}
        \centering
        \includegraphics[width=\linewidth]{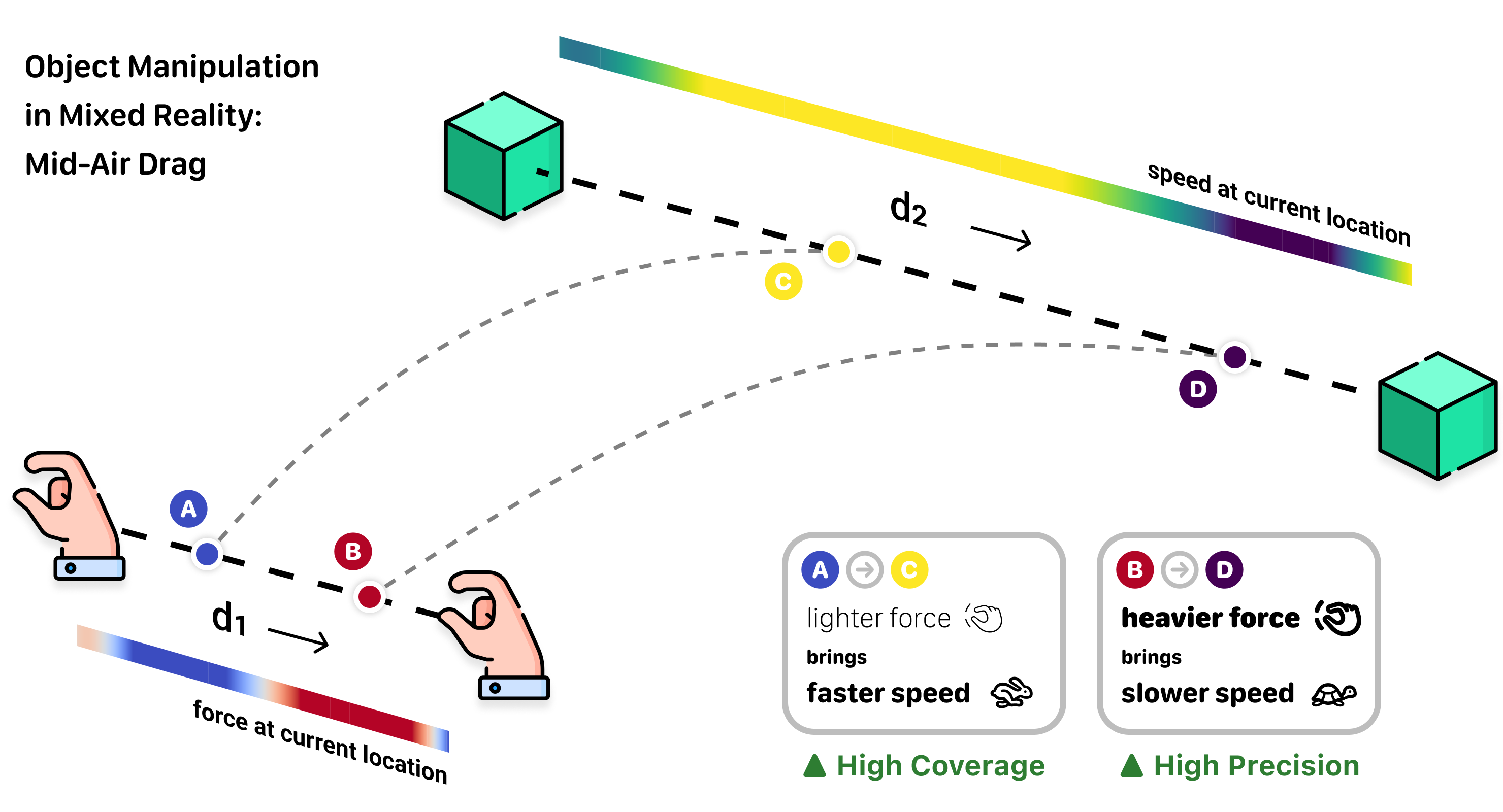}
        \caption{Interaction Process. Users initiate a drag by pinching and holding an object, where applying lighter force (\raisebox{-0.2em}{\includegraphics[height=0.9em]{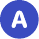}} $\rightarrow$ \raisebox{-0.2em}{\includegraphics[height=0.9em]{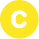}}) results in faster movement for broad coverage, and subsequently applying heavier force (\raisebox{-0.2em}{\includegraphics[height=0.9em]{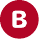}} $\rightarrow$ \raisebox{-0.2em}{\includegraphics[height=0.9em]{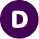}}) slows the movement for precise placement, reinforcing the ``friction'' metaphor—higher force leads to increased friction, reducing speed.}
        \label{fig:force-concept-drag}
    \end{subfigure}
    \hfill
    \begin{subfigure}{0.24\linewidth}
        \centering
        \includegraphics[width=\linewidth]{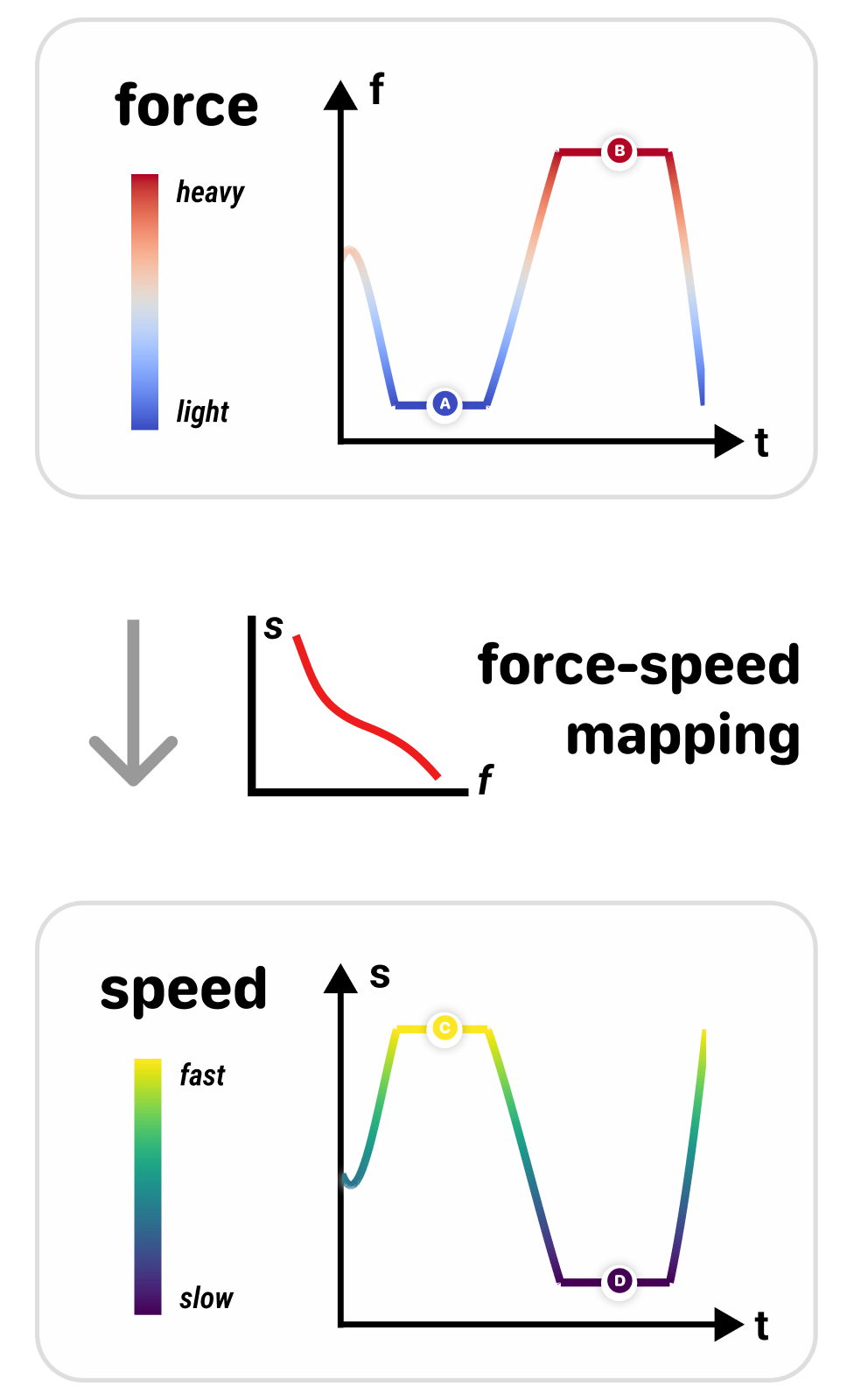}
        \caption{Force-Tracking Speed Mapping. The force values are mapped to tracking speed values via an inverse relationship.}
        \label{fig:force-concept-mapping}
    \end{subfigure}
    \vspace{1.2em}
    \caption{ForcePinch Interaction. This figure illustrates how pinch force modulates object movement speed in a continuous mid-air drag interaction. The force applied at each corresponding spatial position is represented by a color gradient (red = heavy, blue = light), while the corresponding movement speed is visualized in a separate gradient (yellow = fast, purple = slow). ForcePinch allows users to control object translation dynamically, balancing high-speed movement for large-scale adjustments with fine-grained control for precise positioning.}
    \Description{This figure visualizes the ForcePinch interaction. In (a), users modulate tracking speed in mid-air drag by varying pinch force—light force allows fast movement and heavy force slows down movement. A color gradient represents force (red = heavy, blue = light) and speed (yellow = fast, purple = slow). In (b), the mapping from force to tracking speed is shown as an inverse curve.}
    \label{fig:force-concept}
\end{figure*}

\section{ForcePinch}

In this section, we detail the development of ForcePinch, covering our design goals and rationale, software design, and the hardware setup. We conclude with an overview of the system pipeline. The core concept of ForcePinch is illustrated in Figure~\ref{fig:force-concept}.

\subsection{Design Goals}
Our overarching goal is to develop a real-time, intuitive interaction method enabling users to effectively adjust Control-Display (C-D) gain according to their immediate intent. 
To ensure our technique aligns with this objective, we established four design goals (DGs) grounded in established HCI principles and theories. 
These DGs guided the development of ForcePinch:

\textbf{DG1: Enhance Control Granularity.}
Derived from Fitts' Law~\cite{mackenzie1992fitts}, which highlights the inherent tradeoff between movement speed and target precision. Effective 3D interaction requires navigating this tradeoff dynamically. Therefore, a key goal is to provide users with \textit{\textbf{continuous}} control over interaction sensitivity (e.g., C-D gain) to allow for both rapid, coarse movements and slow, precise adjustments within a single, unified interaction modality.

\textbf{DG2: Ensure Intuitive Mapping.}
Based on Norman's Principle of Mapping~\cite{norman2013design}, the relationship between a user's physical actions and the resulting system response must be natural, predictable, and easily understandable. 
An intuitive mapping is crucial as it directly reduces the cognitive effort required to translate user intention into system action, thereby minimizing \textit{the Gulf of Execution}.

\textbf{DG3: Integrate Non-invasive Embodied Interactions.}
Inspired by theories of Embodied Cognition and Interaction~\cite{dourish2001action}, this goal emphasizes utilizing intuitive, physical actions that leverage users' innate understanding derived from interacting with the physical world. 
Crucially, as many spatial interactions are inherently embodied, introducing a new input channel must be done seamlessly. 
Drawing from Cognitive Load Theory~\cite{sweller1988cognitive} and principles of Direct Manipulation~\cite{shneiderman1983direct}, the newly introduced control mechanism must integrate smoothly with the primary task, avoiding explicit mode switches or complex action sequences that could increase cognitive load or disrupt the user's focus.

\textbf{DG4: Provide Continuous and Perceivable Feedback.}
Stemming from Norman's Principle of Feedback~\cite{norman2013design}, users require immediate, continuous, and clearly perceivable information about the system's state and the consequences of their actions. 
Effective feedback allows users to understand the system's response, confirm their actions have the intended effect, and make necessary adjustments.

\subsection{Force and Pinch}

To design an interaction technique addressing our design goals, we drew inspiration from everyday object manipulation, where fingertip force subtly conveys important cues and enables fine motor control.  
Humans instinctively modulate fingertip pressure to adjust friction, enabling precise object control---whether delicately flipping the pages of a book or firmly gripping a glass. Inspired by this intuitive concept of ``friction,'' we propose a direct mapping between the continuously modulated force applied during a pinch gesture and a key interaction parameter, such as the pointer's tracking speed (C-D gain).

The ForcePinch technique combines two fundamental elements: the continuous modulation of force and the discrete pinch gesture. Understanding the inherent characteristics of each reveals why their synergy holds significant potential for meeting our design goals.

\smallskip
\textbf{Force} or pressure, as an input channel, possesses several key characteristics relevant to our interaction design:

\textit{Continuity and High Resolution}: force is an analog, continuous variable. Humans can exert and perceive fine gradations of force~\cite{jones1989matching}. This continuous nature allows for high-resolution control, mapping directly onto continuous system parameters. This intrinsic continuity is fundamental for achieving nuanced control.

\textit{Proprioceptive Feedback}: The exertion of force provides inherent proprioceptive feedback---the body's internal sense of effort and muscle tension~\cite{proske2009kinaesthetic}. Users feel how much force they are applying, even without external system feedback, contributing to a sense of direct connection and control. 

\textit{Intuitive Modulation}: Humans naturally modulate force when interacting with the physical world, often associating increased force with increased control, stability, or precision (e.g., gripping a tool tighter for a delicate task)~\cite{norman2013design}. 
This makes force an intuitive channel for controlling interactions that require careful modulation.

\textit{Eyes-Free Potential}: As force relies on proprioception, it can potentially reduce the need for constant visual monitoring of the control itself, freeing visual attention for the primary task~\cite{zhai1996influence}.

\smallskip
\textbf{The pinch gesture} (typically thumb and index finger opposition) is widely adopted in HCI, particularly in spatial interfaces:

\textit{Established Selection Metaphor}: Pinching is a common and relatively unambiguous gesture used to signal discrete intent, such as selecting an object, initiating a grasp, or starting/stopping a manipulation~\cite{argelaguet2013survey}.
Its meaning is often well-understood by users, reducing the learning curve.

\textit{Compatibility}: The gesture is readily detectable by various tracking systems, including in major commercialized VR/AR headsets.

\smallskip
\textbf{ForcePinch} integrates continuous force sensing within the pinch gesture, combining their strengths to address our design goals:

\textbf{Meeting DG1 (Control Granularity)}:
The continuous nature and high resolution of force input~\cite{ramos2004pressure} is leveraged directly. By modulating force during an active pinch, users can continuously adjust parameters like tracking speed, enabling smooth transitions between speed and precision.

\textbf{Meeting DG2 (Intuitive Mapping)}: 
The intuitive modulation associated with force is key. For example, mapping increasing force to decreasing tracking speed (for precision) aligns with the natural tendency to exert more controlled pressure during fine motor tasks. 
This mapping, applied within the context of the familiar pinch gesture \cite{mine1997moving}, aims to minimize the Gulf of Execution.

\textbf{Meeting DG3 (Embodied Interaction Integration)}: 
ForcePinch builds upon the embodied pinch gesture already used for selection/grasping. Adding force modulation to this existing action avoids introducing separate modes or gestures, thus integrating seamlessly with the primary task.
It enhances the primary interaction rather than replacing it, leveraging the physicality of the pinch itself.

\textbf{Meeting DG4 (Feedback)}: The interaction benefits from multiple feedback layers. The discrete pinch provides clear state feedback (engaged or not). The force input provides inherent proprioceptive feedback \cite{proske2009kinaesthetic}, giving users an internal sense of their control level.

\begin{figure}
    \centering
    \includegraphics[width=\linewidth]{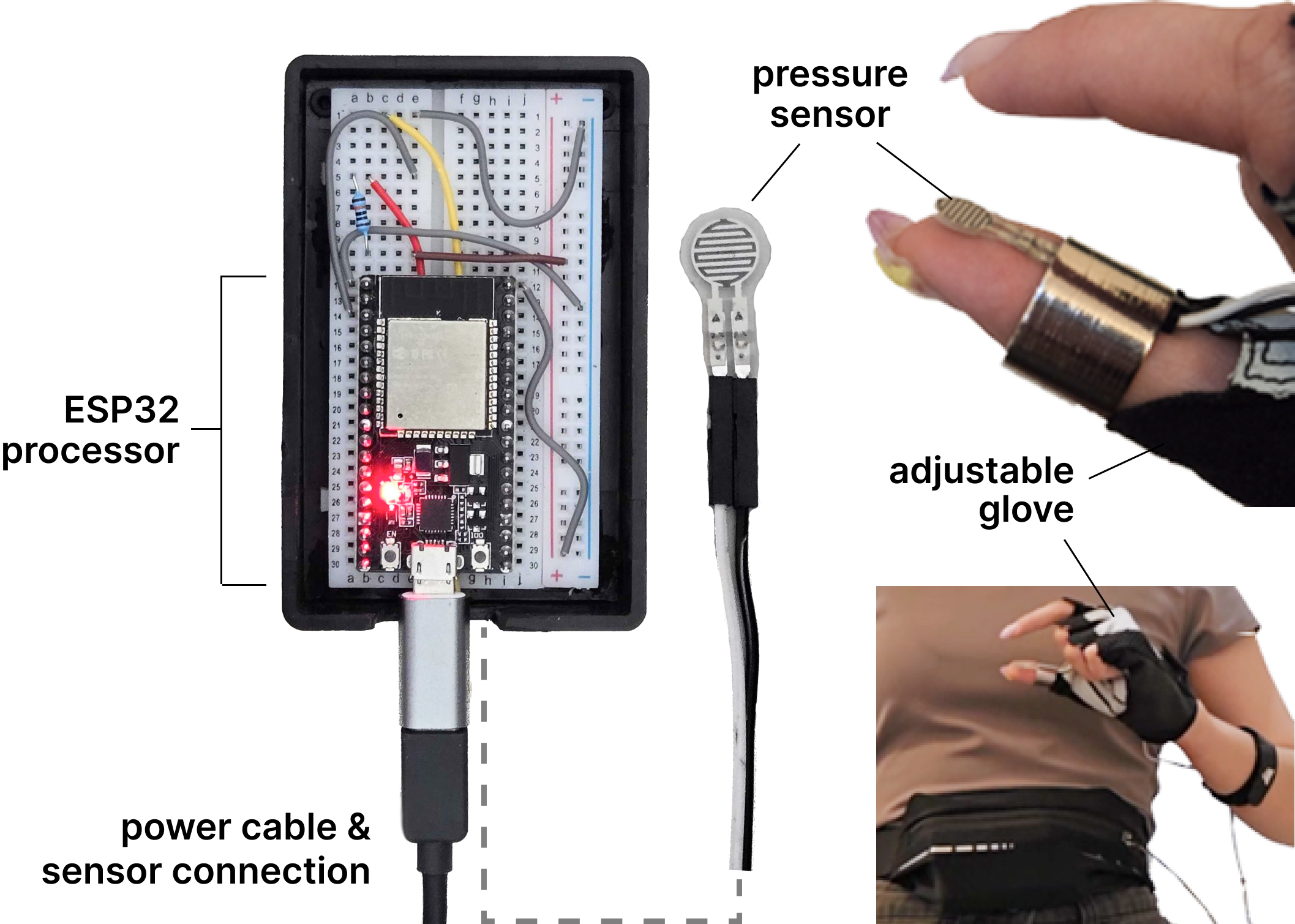}
    \caption{Hardware setup for the force-sensing system. The ESP32 microcontroller, housed in a project box, processes input from a pressure sensor attached to the user's finger via an adjustable glove and ring. The system captures pinch force data and transmits it for interaction control.}
    \Description{This figure depicts the hardware setup. A pressure sensor attached to the user's finger connects to an ESP32 microcontroller housed in a project box, enabling real-time wireless force measurement during interaction.}
    \label{fig:hardware}
\end{figure}

\subsection{Force-Speed Mapping}

The force applied during a pinch gesture offers a natural and expressive input dimension for modulating object movement. To leverage this, we define a mapping between pinch force and tracking speed, allowing users to intuitively control how far an object moves based on how much pressure they exert during a drag.

\revision{
To capture pinch force in real time, we built a lightweight, wireless pressure-sensing system (Figure \ref{fig:hardware}) using a thin-film pressure sensor worn on the finger. The digital signal is processed by an ESP32-based pressure measurement unit and transmitted wirelessly to the headset. The captured force is then mapped to a tracking speed using a continuous Force-Speed Mapping function. Finally, the object’s displacement is computed by scaling the user’s hand movement by this speed, enabling dynamic, force-responsive control.
}

To align with physical intuition, we adopt a ``friction'' metaphor: greater pinch force increases perceived friction, reducing object speed, while lighter force lowers friction, allowing faster movement. This inverse relationship effectively modulates the control-display (CD) ratio—low force results in a high CD ratio for broad, rapid motion, while high force reduces the ratio for fine-grained precision.

This design supports fluid transitions between coarse and fine control within a single gesture. As shown in Figure~\ref{fig:force-concept-drag}, users can move quickly with light force, then slow down with increased pressure for precise positioning—without altering their physical hand speed or switching modes. Releasing the pinch completes the interaction.

\revision{
\subsection{Usability-Driven Design}
To enable smooth and controllable experience, we introduce three design components focused on calibration, feedback, and stability.
\paragraph{Individualized Calibration} 
To account for individual differences in force perception and control, we calibrate each user’s force range by mapping their minimum, moderate, and maximum pinch forces to tracking speeds—fast, neutral, and slow. This ensures personalized, smooth force-to-speed control. We use cubic Hermite spline interpolation to construct the mapping, clamping out-of-range values to avoid instability. Full calibration details are provided in the appendix.
\paragraph{Feedback Design}
We designed a dynamic dot cursor that visually encodes pinch force in real time by continuously changing its size—shrinking as force increases (Figure~\ref{fig:teaser-fast}$\rightarrow$\ref{fig:teaser-slow}). Positioned at the pointer, this metaphor of squeezing a soft object leverages embodied experience, helping users intuitively interpret their applied force and achieve more confident, precise control in speed-sensitive tasks.
\paragraph{Error Handling}
To reduce jitter when releasing objects, we apply a rollback mechanism that reverts the object’s position to the moment of peak force within the previous 0.2 seconds. This compensates for the exaggerated motion caused by sudden force drops and minor hand movements, ensuring smoother release behavior without disrupting the user’s interaction.
}

\section{User Study}


To assess the viability and unique characteristics of the ForcePinch technique, we conducted a controlled comparative user study. 


We compared ForcePinch against a constant-scale baseline and two established techniques serve as critical benchmarks for adaptive tracking speed control: Go-Go~\cite{poupyrev1996go}, which uses distance-based scaling, and PRISM~\cite{frees2007prism}, which adjusts speed based on hand velocity. Unlike these kinematic approaches that tightly couple control to hand movement, ForcePinch enables decoupled, force-based modulation.
The study examined user performance and perceived usability across a series of representative 1D, 2D, and 3D manipulation tasks, aiming to reveal the strengths, limitations, and distinct characteristics of ForcePinch for spatial interaction design in comparison to these established kinematic-based techniques.


\begin{figure}
    \centering
    \includegraphics[width=\linewidth]{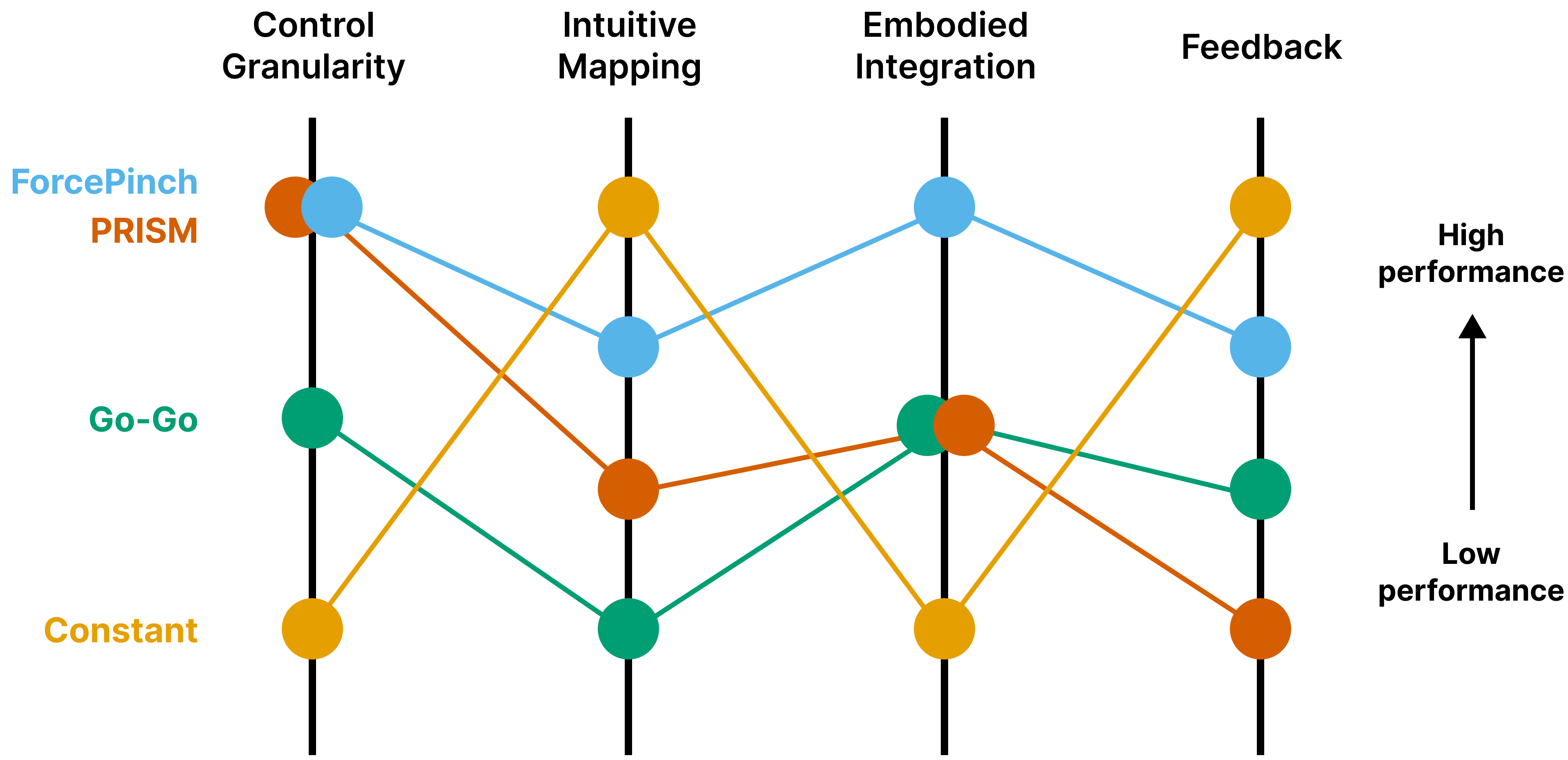}
    \caption{Qualitative ranking of experimental conditions with respect to each design goal. Rankings are relative and illustrative, not derived from precise quantitative metrics.}
    \Description{This figure presents a qualitative comparison of four tracking speed control techniques—ForcePinch, PRISM, Go-Go, and Constant—across four design goals: Control Granularity, Intuitive Mapping, Embodied Integration, and Feedback. The y-axis represents performance (from low to high), and each technique is represented by a colored line (ForcePinch in blue, PRISM in orange, Go-Go in green, Constant in yellow). Lines connect performance levels across the four design goals, showing relative strengths and weaknesses of each method. For example, ForcePinch ranks highest in Control Granularity and Embodied Integration, while PRISM scores lower in Feedback. Constant performs consistently low across all dimensions. These rankings are illustrative, not based on quantitative data.}
    \label{fig:user-study_conditions}
\end{figure}

\subsection{Experimental Conditions}

To conduct a robust comparison, our study included a \Constant{}-scale ($1:c$) baseline method alongside three distinct adaptive tracking speed control techniques: \GoGo{}~\cite{poupyrev1996go}, scaling movement based on hand displacement from an anchor point; \PRISM{}~\cite{frees2007prism}, modulating tracking speed based on hand movement velocity; and our proposed \ForcePinch{}, adjusting tracking sensitivity via pinch force.

We selected these specific methods because they are well-established spatial interaction methods, and they align uniquely with our design goals (DGs), enabling a nuanced evaluation (Figure \ref{fig:user-study_conditions}):

\textbf{DG1 (Control Granularity)}:
ForcePinch offers direct, continuous control over sensitivity via force. 
PRISM provides dynamic granularity \emph{indirectly} through hand velocity. 
Go-Go adapts granularity based on distance, primarily affecting reach rather than offering fine-grained precision control on demand. 
Constant offers no dynamic granularity adjustment.

\textbf{DG2 (Intuitive Mapping)}: 
ForcePinch uses a ``friction'' metaphor (more force $=$ slower speed). 
PRISM relies on the generally intuitive speed-to-speed correlation. 
Go-Go's distance-to-reach mapping, while functional, is arguably less intuitive for deliberate speed modulation. 
Constant's $1:c$ mapping is highly intuitive but inflexible.

\begin{figure}
    \centering
    \includegraphics[width=\linewidth]{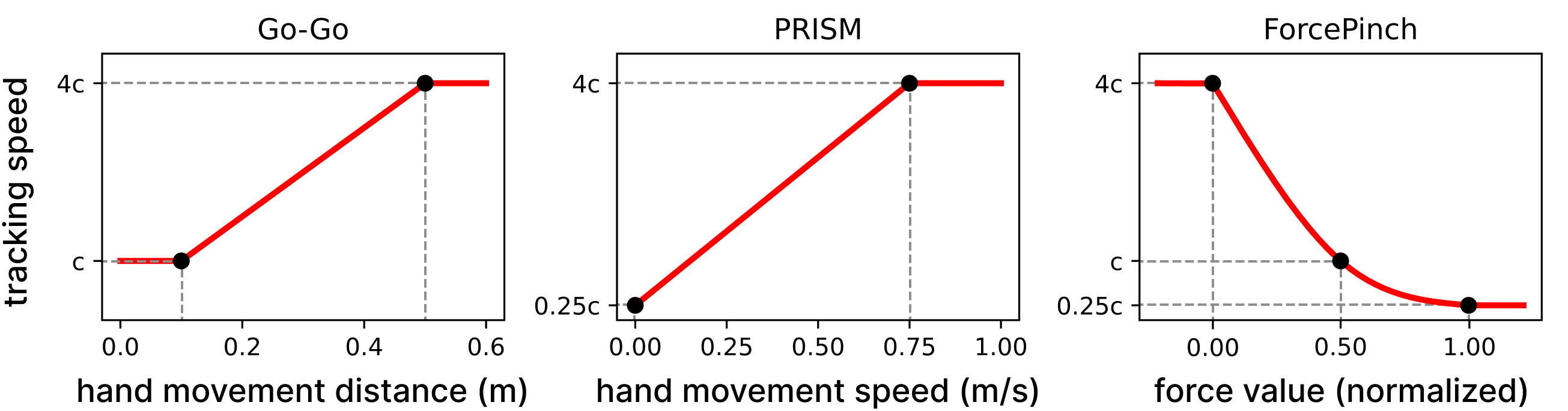}
    \caption{Mapping Functions. Mapping Functions. Tracking speed increases with hand displacement in Go-Go (left), with hand velocity in PRISM (middle), and decreases with pinch force in ForcePinch (right). For 1D and 2D tasks, $c=0.5$; for 3D tasks, $c=1.0$.}
    \Description{This figure shows three line graphs comparing the mapping functions of three adaptive tracking speed techniques: Go-Go (left), PRISM (middle), and ForcePinch (right). In the Go-Go graph, the x-axis represents hand movement distance (in meters), and the y-axis shows tracking speed. The curve starts at speed c for distances under 0.1m and increases linearly, reaching 4c at 0.5m. In the PRISM graph, the x-axis represents hand velocity (in meters per second), and the y-axis shows tracking speed. The curve begins at 0.25c for stationary hands and increases linearly, reaching 4c at a velocity of 0.75 m/s. In the ForcePinch graph, the x-axis shows normalized pinch force values from 0 to 1, and the y-axis shows tracking speed. The curve starts at 4c for minimal force and decreases smoothly to 0.25c at maximum force using a cubic spline. Together, the graphs illustrate how each technique modulates tracking speed: Go-Go based on displacement, PRISM based on velocity, and ForcePinch based on pinch force. For 1D and 2D tasks, c = 0.5; for 3D tasks, c = 1.0.}
    \label{fig:mapping_functions}
\end{figure}

\textbf{DG3 (Embodied Interaction Integration)}: 
ForcePinch integrates force modulation into the existing pinch gesture, crucially decoupling speed control from the primary hand movement without extra modes. 
In contrast, PRISM and Go-Go achieve integration by coupling speed control directly to the kinematics (velocity or distance) of the primary hand movement itself. 
Constant integration is trivial as it adds no control complexity.

\textbf{DG4 (Feedback)}: 
ForcePinch provides inherent proprioceptive feedback through the sensation of applied force. 
Go-Go also has proprioceptive awareness of their hand movement.
PRISM's feedback is mainly the resulting object speed, offering a less direct indication of the current tracking speed compared to ForcePinch's felt force or Go-Go's movement awareness.
All conditions also included visual cursor feedback (e.g., circle size varying based on the current tracking speed).
In Constant, the user will see the cursor in a consistent size.

\begin{figure*}
    \centering
    \begin{subfigure}{0.32\linewidth}
        \centering
        \includegraphics[width=\linewidth]{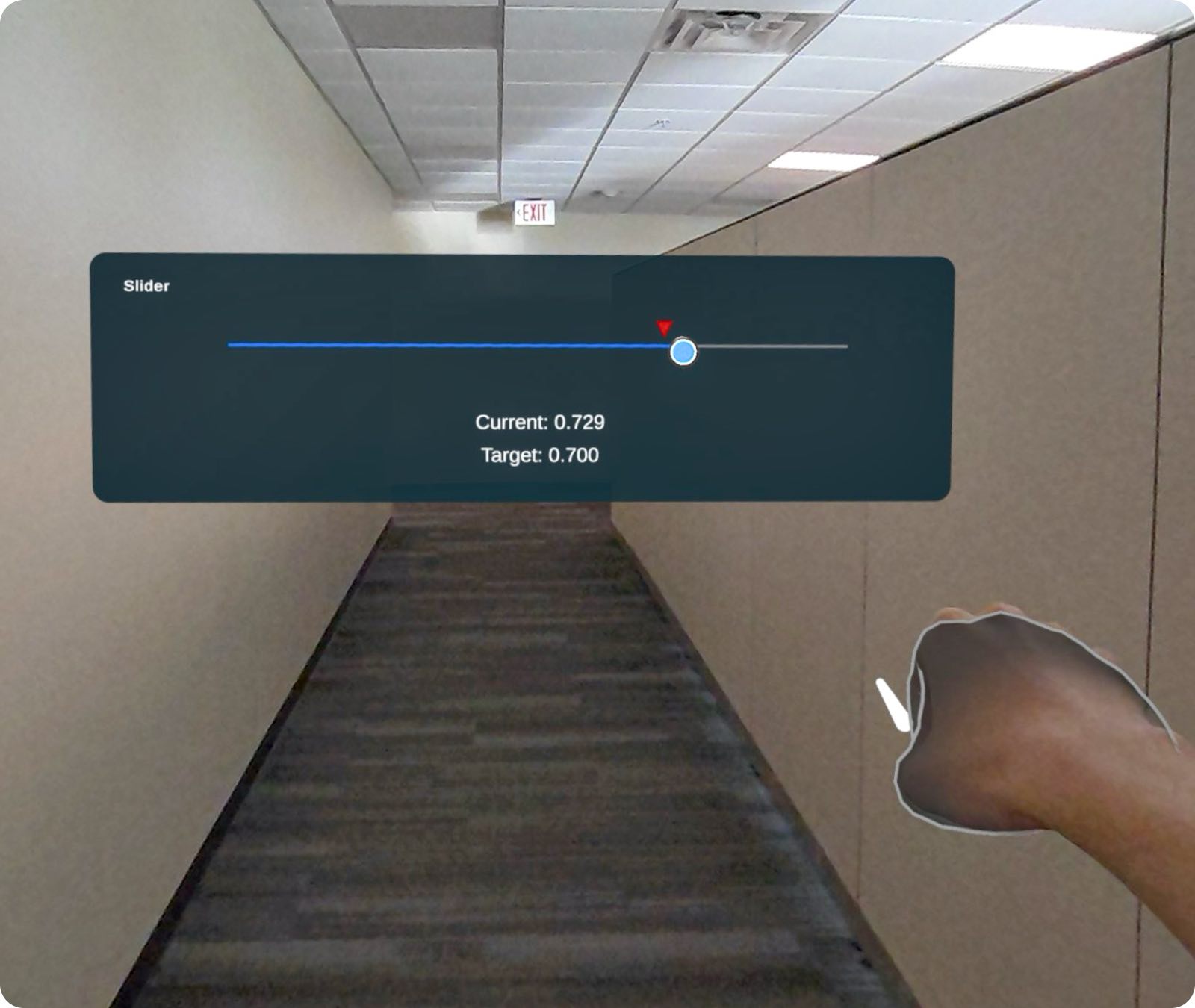}
        \caption{Task 1: 1D-Slider.}
        \label{fig:user-study-scene-1}
    \end{subfigure}
    \hfill
    \begin{subfigure}{0.32\linewidth}
        \centering
        \includegraphics[width=\linewidth]{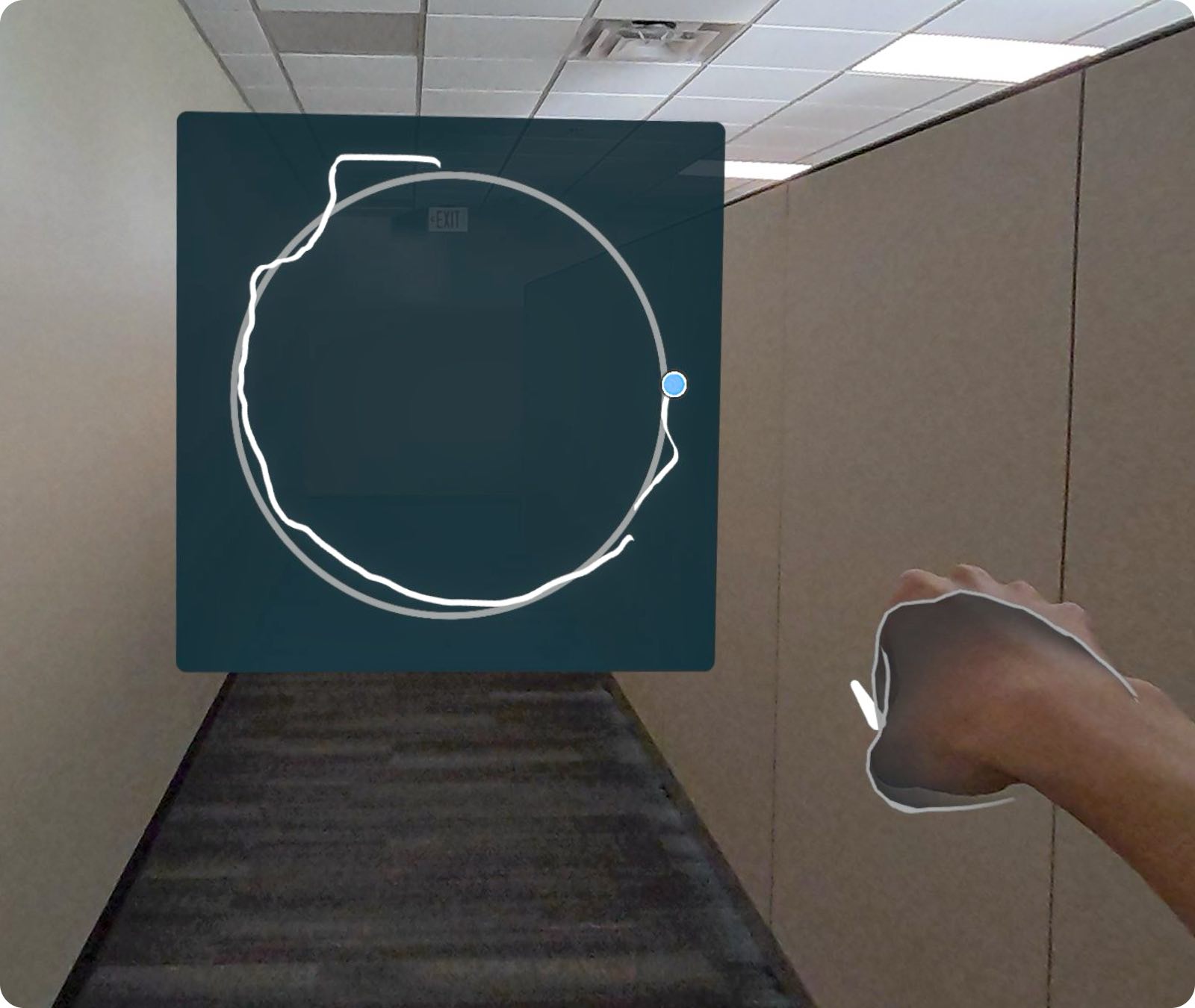}
        \caption{Task 2: 2D-Tracing.}
        \label{fig:user-study-scene-2}
    \end{subfigure}
    \hfill
    \begin{subfigure}{0.32\linewidth}
        \centering
        \includegraphics[width=\linewidth]{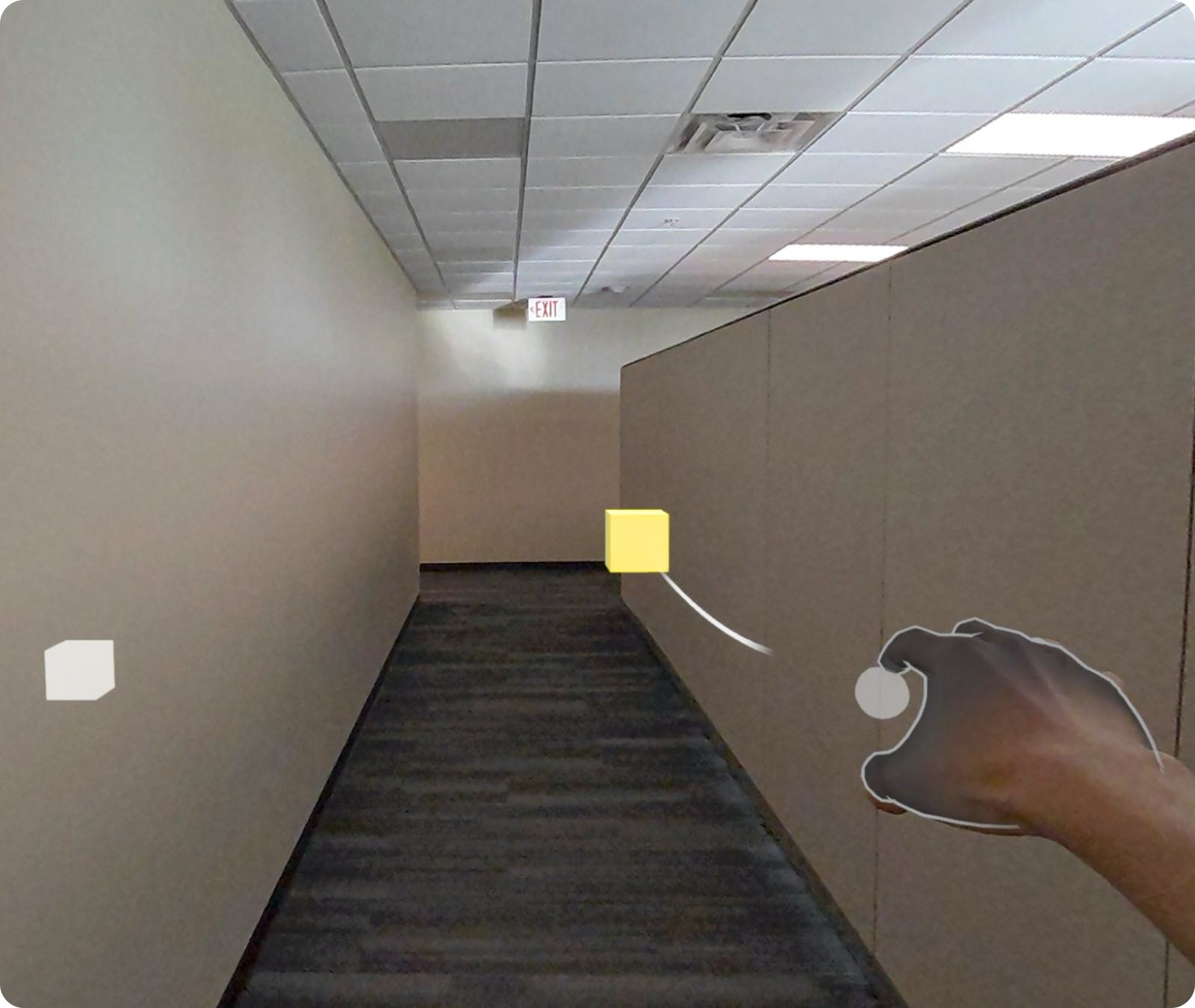}
        \caption{Task 3: 3D-Placement.}
        \label{fig:user-study-scene-3}
    \end{subfigure}
    \vspace{1.5em}
    \caption{Participant's headset view during the user study tasks. A mixed reality scene was used to allow direct visibility of the user's hands, enhancing their perception of hand movements during interaction.}
    \Description{This figure shows the user's view during the three user study tasks. (a) 1D slider task, (b) 2D shape tracing on a canvas, and (c) 3D cube placement, all rendered with passthrough for better hand visibility.}
    \label{fig:user-study-scene}
\end{figure*}

Through pilot studies with six participants, we determined the parameters for four techniques (conditions). The resulting tracking speed functions are shown in Figure~\ref{fig:mapping_functions}. The specific settings for each condition are as follows:
\begin{itemize}[leftmargin=*]
    \item \textbf{\Constant{}:} A constant control-display (C-D) ratio baseline ($1:c$). To better balance efficiency and accuracy, we choose a slower tracking speed $c=0.5$ for 1D and 2D tasks to account for their smaller movement ranges, and a larger tracking speed $c=1.0$ for 3D tasks to facilitate long-distance spatial movement.
    \item \textbf{\GoGo{}:} A distance-based scaling method where tracking speed increases with displacement. An initial constant-speed phase is applied for displacements below 0.1m to prevent excessively slow movements. During this phase, tracking speed is set to $c$, matching the \Constant{} condition. Beyond 0.1m, tracking speed increases linearly with distance, reaching a maximum of $4c$ at 0.5m. 
    \item \textbf{\PRISM{}:} A velocity-based scaling method where tracking speed increases with hand movement velocity. The tracking speed ranges from $0.25c$ at 0 m/s to a maximum of $4c$ at 0.75 m/s. To maintain controllability, hand velocity above 0.75 m/s do not result in further increases in tracking speed.
    \item \textbf{\ForcePinch{}:} A force-based scaling method. Each participant’s pinch force range is normalized to (0, 1), and a cubic spline interpolation defines the mapping with three key points: (0, 4$c$), (0.5, $c$), and (1.0, 0.25$c$). This inverse mapping (higher force $\rightarrow$ lower speed) mimics the sensation of ``friction.''
\end{itemize}

To ensure consistency across conditions, all mappings share the same $c$ value and attempt to maintain similar lower and upper bounds of tracking speed for the same task. However, for \GoGo{}, which only allows one-directional scaling, it is not possible to match the lower bound with those of \PRISM{} and \ForcePinch{}, and here we set it's lower bound the $c$, same as \Constant{} condition.

\subsection{Tasks}

Participants completed three tasks designed to evaluate control precision and usability across 1D, 2D, and 3D interactions, as presented in Figure \ref{fig:user-study-scene}:

\textbf{T1: 1D Slider.}
A virtual slider was placed 2 meters in front of the user at a height of 1.5 meters, with a total length of 1.5 meters. Participants used ray casting to select the slider handle and then dragged it horizontally to match a target position. Both the current and target values were displayed beneath the slider with three-decimal-point precision. Each participant completed one training trial and five testing trials, each with a randomly assigned target value. Target values were selected to fall within a range of 0.5 to 0.8, ensuring sufficient movement to differentiate technique performance.

\textbf{T2: 2D Tracing.}
Participants used ray casting to set the starting point and traced shapes displayed on a 1m × 1m virtual canvas positioned 2 meters away. Each shape was centrally placed with a 10\% padding from all canvas edges. Five distinct shapes were used, and their order was counterbalanced across participants using a 5×5 Latin square. Each participant completed one tracing trial per shape, and they can draw multiple strokes in one trial.

\textbf{T3: 3D Placement.}
Participants used ray casting to select and used hand movement to manipulate a virtual cube in 3D space to align it with a semi-transparent target cube at varying positions. Successful placement required visually matching the object with the target position. Each participant completed one training trial and five testing trials, with targets distributed across a diverse spatial range to encourage full 3D movement. In each trial, the target cube was placed 3-4 meters away from the final target position, requiring substantial repositioning across all three axes.

\subsection{Experimental Setup}

The experiment was conducted using a Meta Quest 3 headset, enabling head-mounted spatial interaction and hand tracking. For the ForcePinch condition, a thin-film pressure sensor was integrated and connected to an ESP32 microcontroller, which wirelessly streamed real-time pressure data to the headset. The sensor was affixed to the thumb pad and remained attached across all four conditions to ensure consistency.
The study was conducted in a clear, obstacle-free physical space measuring approximately $10 \times 10$ feet. Participants used the headset in wireless mode to allow free movement without cable interference.

The virtual environment was developed in Unity, utilizing mixed reality passthrough to allow participants to see their real hands. We employed the Meta XR Interaction SDK to enable hand tracking and object manipulation. To support all four control techniques, we extended the SDK's built-in Distance Grab Interactable component by implementing a custom Movement Provider. Each technique was implemented as a selectable mode within a unified interaction framework to ensure consistency across conditions.



\subsection{Participants}

Twenty participants (9 female, 11 male), all university students aged 20 to 29, were recruited from local campuses. Participants had varied levels of prior VR experience: 2 were frequent users, 1 used VR regularly (a few times per month), 4 used it occasionally (a few times per year), 11 had tried VR once or twice, and 2 had no prior VR experience. Each participant received a \$15 gift card as compensation.


\subsection{Design and Procedures}

The study aimed to evaluate the effectiveness of our tracking speed control techniques in supporting task performance, while also examining their cognitive and physical demands. Specifically, we sought to understand whether increased responsiveness improves user experience and which method performs best across interaction tasks.

We employed a within-subject design in which all participants experienced all four control conditions. To mitigate order effects, condition sequences were counterbalanced using a Latin square (4 groups). The task order was fixed (T1 $\rightarrow$ T2 $\rightarrow$ T3) to reflect increasing interaction dimensionality.
At the start of the session, participants were briefed on the study goals and introduced to the four tracking speed control techniques and the three tasks. The researcher provided a live demonstration of the full procedure. Participants then completed a pressure sensor calibration and were equipped with the sensor and a waist-mounted bag containing the ESP32 hardware.
For each combination of task and control condition, participants were given time to freely explore and train with the method before proceeding to five formal trials. \revision{Since the focus of this study was not on learnability, and the complexity of exploring each method varied depending on the task, we did not enforce uniform training durations. However, to prevent overtraining, each training session was limited to a maximum of two minutes. After completing each control condition, participants completed the NASA Task Load Index (TLX)~\cite{hart2006nasa}, using a 0–20 scale for each dimension, and provided an overall rating for each method on a 1–7 scale to assess general experience.} In addition, semi-structured interviews were conducted after each task to collect qualitative feedback on the user experience for each technique. The whole study procedure lasts for 90 minutes.



\subsection{Measures}

For each trial, we defined the interaction process as the interval beginning with the initial selection and concluding with the final deselection of the target object. During each trial, we recorded data at a frequency of 100 Hz (every 0.01 seconds), including timestamps, positions and selection states of the target object, hand positions, and tracking speed. We then processed this raw data to calculate the following performance measures for each trial: \textit{operation time}, \textit{error distance}, \textit{number of operations}, and the \textit{travel distances} of both the hand and the object. The error distance was computed as the Euclidean distance between the object's final position and the target position for Tasks 1 and 3, and as the Euclidean distance between sampled points and their nearest points on the target path for Task 2. Additionally, we rescaled TLX responses proportionally from their original range (0–20) to a 1–7 scale, where higher scores indicate better outcomes.

\subsection{Statistical Analysis}

For dependent variables (or their transformed counterparts) satisfying the assumption of normality, we applied linear mixed-effects modeling (LMM) to evaluate the effects of independent variables on these measures \cite{bates2015fitting}. \revision{Following standard practice, we applied a log transformation to time data to normalize the distribution. This allowed us to apply LMM appropriately.} Linear mixed-effects modeling was chosen due to its advantage over repeated-measures ANOVA, particularly its robustness to violations of sphericity \cite{field2012discovering}. 
For dependent variables exhibiting non-normal distributions, we employed generalized linear mixed-effects models (GLMMs)~\cite{bolker2009generalized} with appropriate distributional assumptions: a Gamma distribution with a log link for continuous skewed data (e.g., \textit{error distance}) and a Poisson distribution with a log link for count data (e.g., \textit{number of operations}). 
All independent variables (including technique and shape for Task 2) and their interactions were modeled as fixed effects, with participant intercepts included as random effects to account for within-subject variability. We assessed the significance of fixed effects using likelihood ratio tests. Pairwise comparisons were conducted using Tukey’s Honest Significant Difference (HSD) test on the estimated marginal means (EMMs) obtained from the fitted models \cite{lenth2016least}. Diagnostic checks for model assumptions—specifically homoscedasticity and normality of residuals—were visually conducted using predicted-versus-residual and Q–Q plots for models assuming normality.
Statistical significance is reported using asterisks as follows: \(p < .05 (*)\), \(p < .01 (**)\), and \(p < .001 (***)\). Additionally, we report means and 95\% confidence intervals (CIs) as indicators of effect size for significant findings. 
\revision{Full statistical details are included in the supplementary file.}

\begin{figure*}
    \centering
    \includegraphics[width=\linewidth]{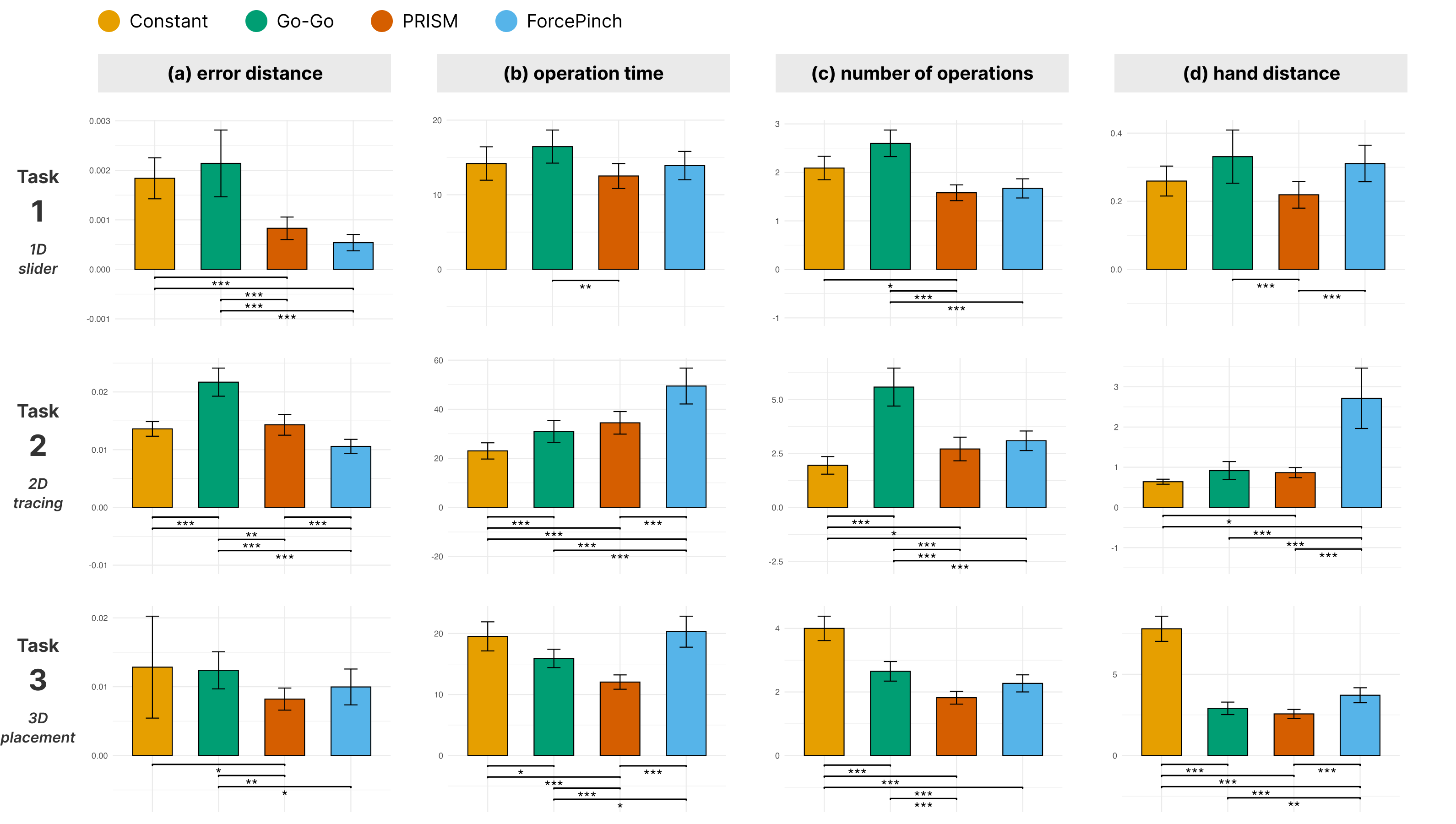}
    \caption{Performance comparison of the four interaction techniques across the three tasks. Each row corresponds to one task, showing four key performance metrics: (a) error distance to the final target (or medium sampling error along the path in Task 2), (b) operation time per trial, (c) number of operations per trial, and (d) total hand travel distance. Bars represent mean values with 95\% confidence intervals. Significance annotations: $***$ for $p<0.001$, $**$ for $0.001 \leq p < 0.01$, and $*$ for $0.01 \leq p < 0.05$.}
    \Description{This figure compares four interaction techniques across all tasks using four metrics: error distance, operation time, number of operations, and hand travel distance. Each row corresponds to a task; bars show mean values with 95\% confidence intervals and statistical significance markers.}
    \label{fig:result-all}
\end{figure*}

\begin{figure*}[htbp]
    \centering
    \includegraphics[width=\linewidth]{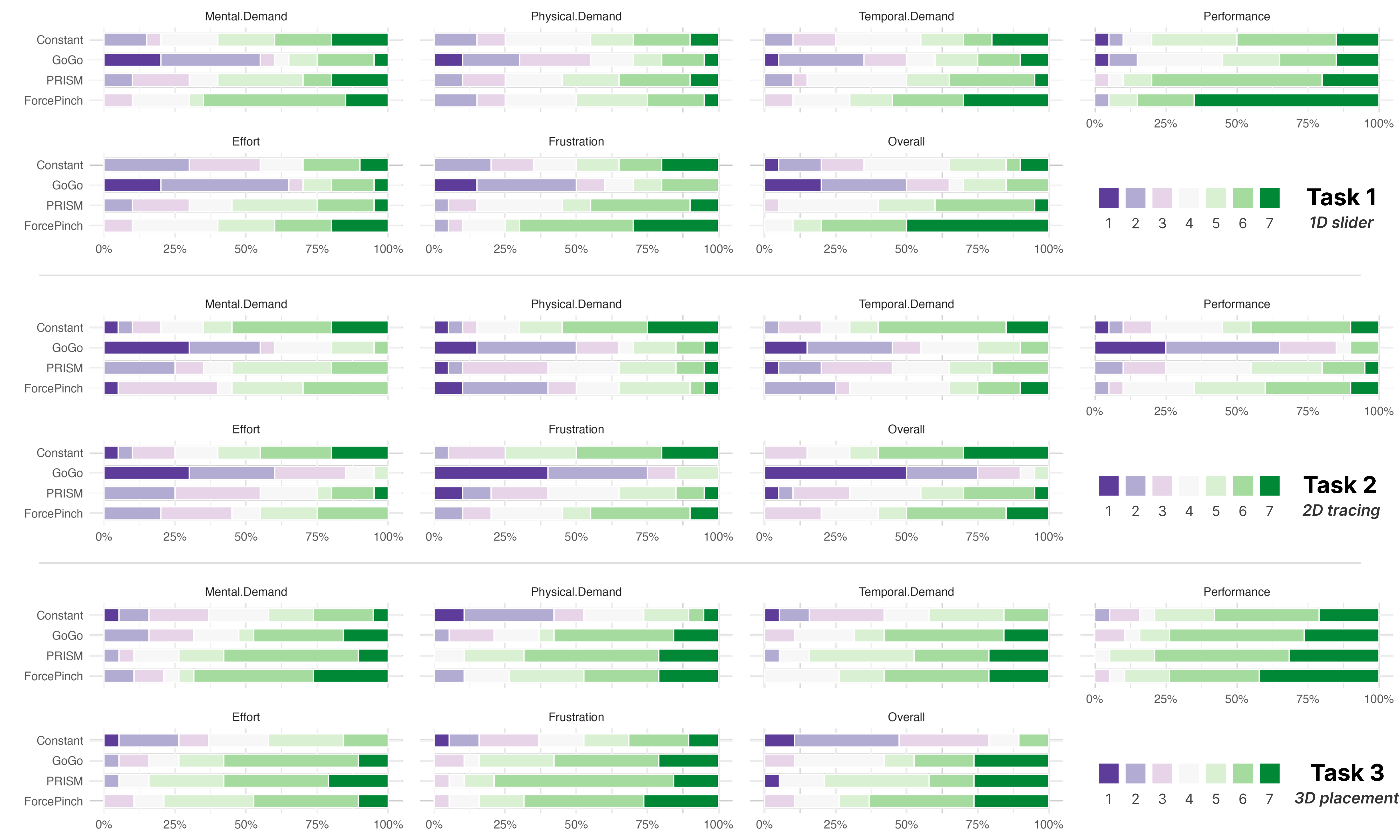}
    \caption{\revision{Subjective ratings for 3 tasks (1D slider, 2D tracing, and 3D placement) based on a 7-point scale, where 1 indicates the most negative and 7 the most positive response. Raw score distributions and per-participant data are included in the supplementary material.}}
    \Description{This figure displays subjective ratings for each of the three tasks using a 7-point scale (1 = negative, 7 = positive). Metrics include mental, physical, and temporal demand, performance, effort, and frustration.}
    \label{fig:tlx-all}
\end{figure*}

\section{Results}

In this section, we present statistical analyses comparing different interaction techniques across three tasks. Results are presented sequentially, moving from the 1D slider to the more complex 3D placement scenarios, highlighting technique characteristics in each context. Complete statistical analyses and anonymized participant data are provided in the supplementary material.

\subsection{Task 1: 1D-Slider}

Statistical results for the four techniques in Task 1 are presented in the first row of Figure~\ref{fig:result-all} and Figure~\ref{fig:tlx-all}.

\topic{Objective Performance.}
We found a significant effect of interaction technique on all performance metrics in Figure~\ref{fig:result-all}. For \textit{error distance}, \ForcePinch{} (0.00054, CI=0.00017) and \PRISM{} (0.00083, CI=0.00023) showed comparable accuracy, both significantly outperforming \Constant{} (0.00184, CI=0.00041) and \GoGo{} (0.00214, CI=0.00067) (all $***$). 
For \textit{operation time}, \ForcePinch{} (13.91, CI=1.89) did not differ significantly from the other techniques, while \PRISM{} (12.51, CI=1.67) was significantly faster than \GoGo{} (16.46, CI=2.22). 
Regarding \textit{number of operations}, \ForcePinch{} (1.67, CI=0.20) was statistically similar to \PRISM{} (1.58, CI=0.16) and \Constant{} (2.09, CI=0.24), but significantly outperformed \GoGo{} (2.60, CI=0.27) ($***$). \PRISM{} also required fewer operations than \Constant{} ($*$) and \GoGo{} ($***$).
For \textit{hand distance}, \ForcePinch{} (0.31, CI=0.05) and \GoGo{} (0.33, CI=0.08) involved significantly more hand movement than \PRISM{} (0.22, CI=0.04) (both $***$).


\topic{Subjective Ratings.}
Subjective ratings are summarized in Figure~\ref{fig:tlx-all} Task 1. Significant effects were found for all dimensions except \textit{physical demand} ($p=0.112$): \textit{mental demand} ($***$), \textit{temporal demand} ($***$), \textit{performance} ($***$), \textit{effort} ($***$), and \textit{frustration} ($***$). 
\ForcePinch{} was rated as significantly less mentally demanding, less effortful, and less frustrating than \GoGo{} (all $***$). It was also significantly preferred over \Constant{} and \GoGo{} ($***$), and over \PRISM{} ($**$). Notably, 70\% of participants ranked \ForcePinch{} as their top choice.

\subsection{Task 2: 2D-Tracing}

Statistical results for the four techniques in Task 2 are presented in the second row of Figure~\ref{fig:result-all} and Figure~\ref{fig:tlx-all}.

\topic{Objective Performance.}
Interaction technique had a significant effect on all performance metrics ($***$; see Figure~\ref{fig:result-all}). For \textit{error distance}, \ForcePinch{} (0.0106, CI=0.0012) significantly outperformed \GoGo{} (0.0217, CI=0.0024), \PRISM{} (0.0143, CI=0.0018), and \Constant{} (0.0136, CI=0.0013) (all $**$ or $***$), demonstrating superior precision in 2D path tracing.
However, \ForcePinch{} (49.47, CI=7.31) required significantly more \textit{operation time} than \Constant{} (23.03, CI=3.31), \GoGo{} (30.97, CI=4.43), and \PRISM{} (34.47, CI=4.59) (all $***$), suggesting that participants spent more time to achieve higher accuracy with the force-responsive method.
In terms of \textit{number of operations}, \ForcePinch{} (3.09, CI=0.45) was comparable to \PRISM{} (2.71, CI=0.55), and outperformed \GoGo{} (5.58, CI=0.88) ($***$). \Constant{} (1.95, CI=0.41) required significantly fewer operations than \ForcePinch{} ($*$), indicating more efficient stroke completion.
For \textit{hand distance}, \ForcePinch{} (2.71, CI=0.75) resulted in significantly more hand movement than all other techniques (all $***$).


\topic{Subjective Ratings.}
As shown in Figure~\ref{fig:tlx-all} Task 2, significant differences were found across all subjective measures (all $***$). Compared to \GoGo{}, \ForcePinch{} was rated as less mentally and temporally demanding, and yielded better perceived performance, lower effort, and lower frustration (all $*/**/***$). 
\ForcePinch{} and \PRISM{} did not differ significantly across any subjective dimension. \Constant{} was rated as having lower \textit{physical demand} ($**$) and \textit{temporal demand} ($*$) than \ForcePinch{}. Overall, \ForcePinch{} was rated similarly to \Constant{} and \PRISM{}, and significantly better than \GoGo{} ($***$).

\subsection{Task 3: 3D-Placement}

Statistical results for the four techniques in Task 3 are shown in the second row of Figure~\ref{fig:result-all} and Figure~\ref{fig:tlx-all}.

\topic{Objective Performance.}
Interaction technique had a significant effect on all performance metrics ($***$; see Figure~\ref{fig:result-all}). For \textit{error distance}, \ForcePinch{} (0.00996, CI=0.00261) outperformed \GoGo{} (0.01238, CI=0.00270) ($*$), while \PRISM{} (0.00820, CI=0.00160) showed significantly better accuracy than both \GoGo{} ($**$) and \Constant{} (0.01284, CI=0.00740) ($*$).
In terms of \textit{number of operations}, \ForcePinch{} (2.27, CI=0.27) reduced the number of required actions by 44.3\% compared to \Constant{} (4.00, CI=0.38) ($***$). \PRISM{} (1.82, CI=0.20) and \GoGo{} (2.65, CI=0.31) also significantly reduced operation counts compared to \Constant{} (both $***$). This result is expected, as the \Constant{} technique offers only limited movement range within arm’s reach, often requiring multiple grabs to complete longer movements.
Surprisingly, \ForcePinch{} (20.30, CI=2.54) required significantly more \textit{operation time} than \PRISM{} (12.04, CI=1.17) ($***$) and \GoGo{} (15.92, CI=1.50) ($*$), with no significant difference from \Constant{} (19.53, CI=2.39).
For \textit{hand distance}, \ForcePinch{} (3.71, CI=0.50) reduced the total hand movement by 52.4\% compared to \Constant{} (7.80, CI=0.77) ($***$), though still more than \GoGo{} (2.90, CI=0.38) ($**$) and \PRISM{} (2.56, CI=0.28) ($***$).


\topic{Subjective Ratings.}
As shown in Figure~\ref{fig:tlx-all} Task 3, subjective differences in user experience were less pronounced than in the previous tasks. Significant effects were observed for \textit{mental demand} ($*$), \textit{physical demand} ($***$), \textit{temporal demand} ($***$), \textit{effort} ($***$), and \textit{frustration} ($**$), but not for \textit{performance} ($p=0.092$).
\ForcePinch{} did not differ significantly from \GoGo{} or \PRISM{} on any subjective metrics, and all three techniques generally received positive feedback. However, \ForcePinch{} outperformed \Constant{} in \textit{mental demand} ($*$), \textit{physical demand} ($***$), \textit{temporal demand} ($***$), \textit{effort} ($**$), and \textit{frustration} ($*$). 
Overall, participants expressed no clear preference among \ForcePinch{}, \GoGo{}, and \PRISM{}, but all three were rated significantly more favorably than \Constant{} ($***$).

\section{Discussion: User Study Key Findings}

The user study evaluated ForcePinch against established baseline (Constant) and adaptive (Go-Go, PRISM) techniques across tasks of varying dimensionality. 
Our results revealed several key trends: ForcePinch demonstrated notable advantages in precision and user preference, particularly in 1D and 2D tasks, though often at the cost of increased completion time and with reduced effectiveness in 3D placement. 
Furthermore, the adaptive techniques generally outperformed the Constant baseline, highlighting the benefits of responsiveness. 
This section delves into these findings to understand their implications.

\subsection{Do we need responsive interactions?}
\topic{Responsive techniques help tasks requiring both rapid movement and fine control.}
Indeed, our results show that responsive interaction techniques (like ForcePinch, Go-Go, PRISM), which adjust control in real time, are generally better than a fixed Constant (1:$c$) control for some VR tasks (Tasks T1, T3). We found responsive methods were often more efficient, needing fewer actions and less hand movement, especially when tasks involved reaching or varying precision demands (Tasks T1, T3). 
Users also preferred them over the Constant control method in the 3D task, finding them less tiring and more efficient. This matches past research~\cite{bowman1997evaluation} showing the limits of basic 1:1 interaction for distant objects and the benefits of adaptive methods in VR~\cite{bowman1999testbed}.

\topic{Constant control benefits tasks needing steady, continuous movement.}
However, responsive control was not always the best choice. For tasks requiring steady and predictable continuous movement, like our 2D tracing (T2), the simple Constant method was actually faster.
We anticipated that for tasks emphasizing continuous, predictable control without significant scale changes, the simplicity of 1:1 mapping can be advantageous.
Responsive techniques often incurred extra costs, such as potentially longer completion times when users leverage fine-grained control for accuracy.
Even with these trade-offs, 
common spatial interactions involving reaching across distances necessitate dynamic shifts between speed and precision, which shows that responsive interaction is valuable and thus justifies the development of techniques like ForcePinch.

\subsection{Is ForcePinch a good responsive method?}

\begin{figure}
    \centering
    \includegraphics[width=\linewidth]{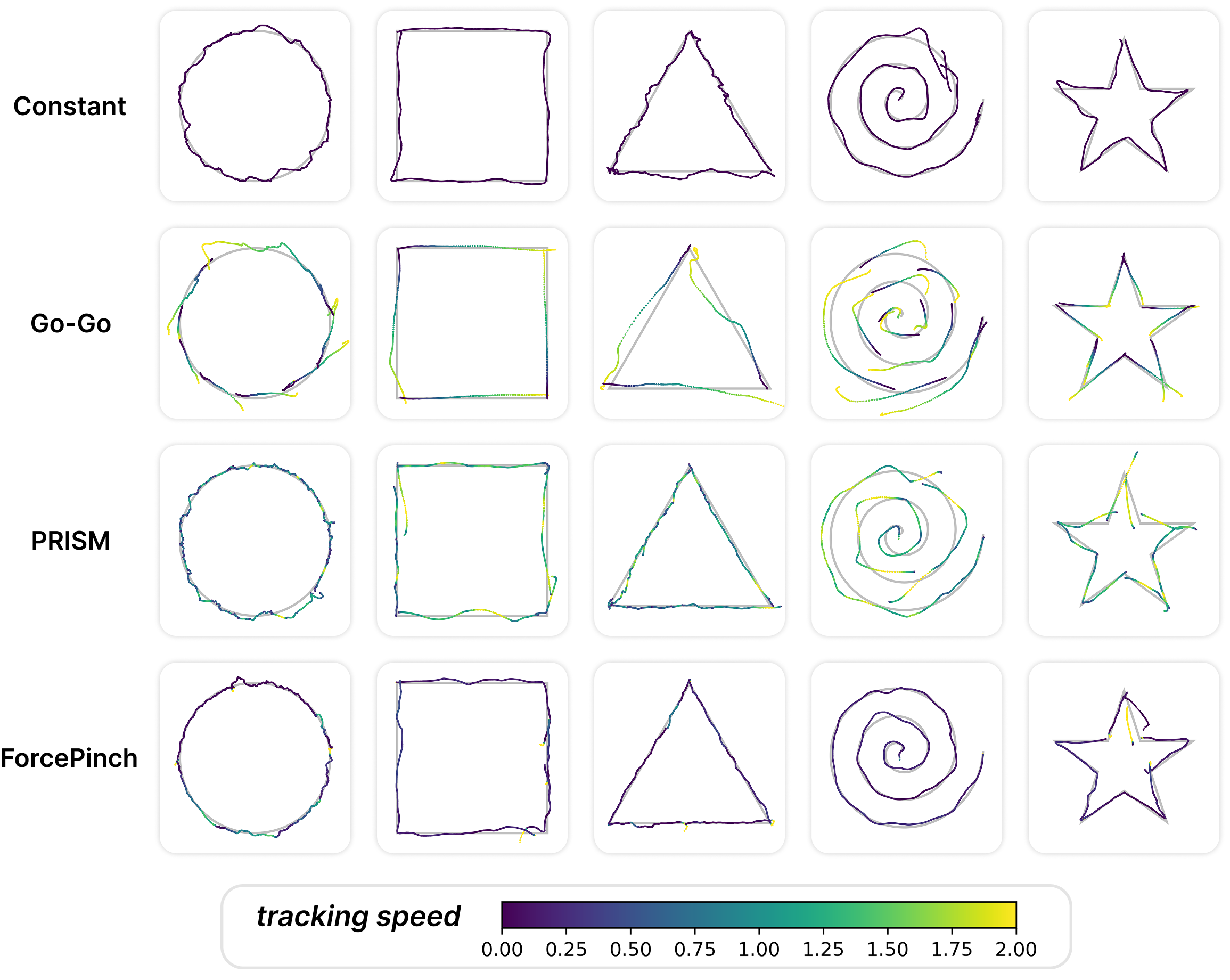}
    \caption{Representative drawings of each shape evaluated in Task 2 across all tracking methods. The shapes, from left to right, are: circle, square, triangle, spiral, and star.}
    \Description{This figure shows user drawings of five shapes (circle, square, triangle, spiral, star) using each tracking technique. ForcePinch lines are generally smoother and closer to targets, indicating improved precision.}
    \label{fig:task2-drawings}
\end{figure}

\topic{ForcePinch intuitively enhances precision.}
Our user study results suggest that ForcePinch improves operation precision (lower error distance) in both 1D and 2D tasks and performs comparably to other techniques in 3D scenarios. This indicates that ForcePinch is effective in enhancing interaction accuracy. While PRISM achieved similar precision in Tasks 1 and 3, ForcePinch outperformed it in Task 2. As shown in Figure \ref{fig:task2-drawings}, two participants' trajectories in Task 2 demonstrate that ForcePinch results in fewer deviations and smoother strokes.
Furthermore, semi-structured interview data support these findings: over half of the participants reported that ForcePinch felt particularly intuitive in 1D and 2D contexts. One participant shared, ``When I want to move quickly, I just pinch lightly. When I need to slow down for fine control, I press harder. The transition feels very natural and efficient.''

\begin{figure}
    \centering
    \includegraphics[width=\linewidth]{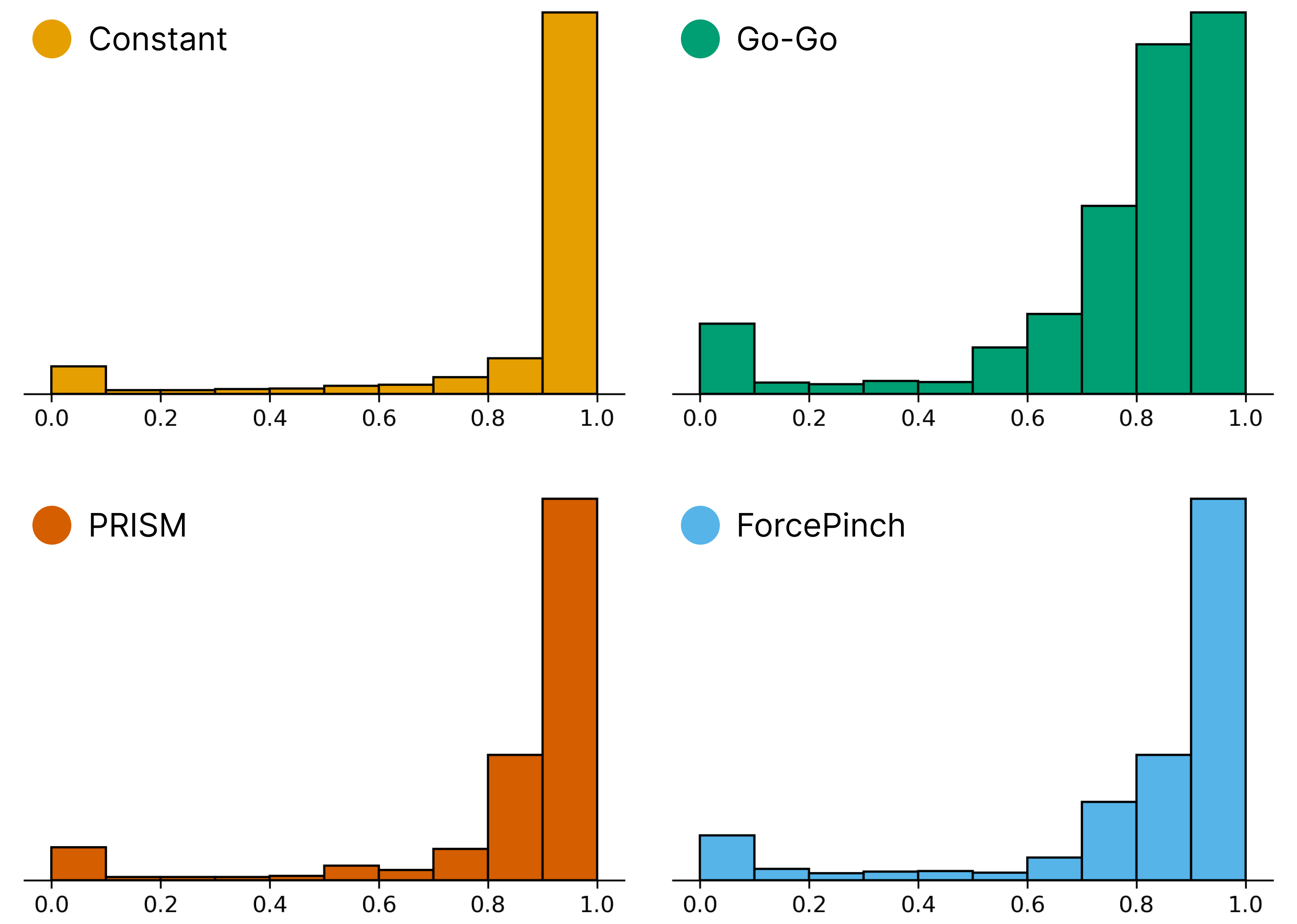}
    \caption{Histogram of time spent at different positions along the slider path, where \textit{x = 0} represents the starting point and \textit{x = 1} represents the target position. Taller bars indicate longer time spent at corresponding positions. ForcePinch enables users to concentrate interaction near the target while maintaining a smooth distribution of time across the path, demonstrating a balance between rapid and precise control.}
    \Description{This figure is a histogram showing how long users spent at different positions along the 1D slider. ForcePinch results in more time spent near the target, reflecting precise final adjustments.}
    \label{fig:slider-hist}
\end{figure}

\topic{ForcePinch does not reduce hand movement distance or task completion time.}
To achieve higher precision, users often apply sustained force to reduce tracking speed. Meanwhile, this also means that covering the same distance requires more hand movement. In our tasks—designed with accuracy as the primary objective—ForcePinch did not reduce the physical distance of hand movement. Similarly, users tended to spend more time on fine adjustments using ForcePinch, leading to longer task completion times.
Figure \ref{fig:slider-hist} presents a histogram of time spent at each slider position during Task 1. The data show that with ForcePinch, users spent more time in regions closer to the target value (1.0), indicating finer control near the goal. Notably, the ForcePinch histogram is also the smoothest among all techniques, reflecting more natural hand movement patterns. We believe that longer hand movement distances are acceptable up to a certain extent, especially for scenarios requiring higher accuracy.

\subsection{How did users adjust tracking speed?}

\begin{figure}
    \centering
    \includegraphics[width=\linewidth]{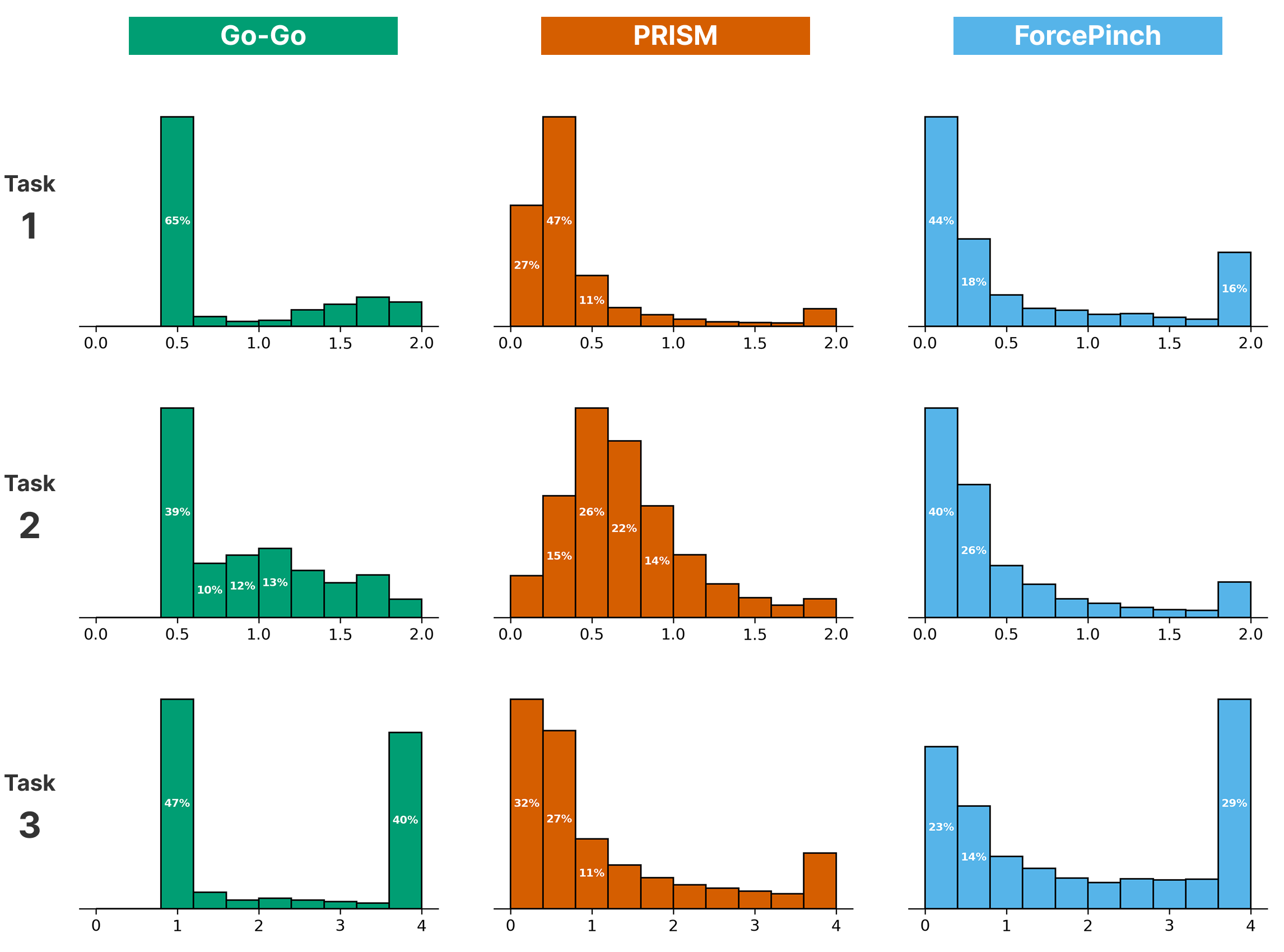}
    \caption{Tracking speed histograms for Tasks 1–3 using Go-Go (left), PRISM (middle), and ForcePinch (right). ForcePinch demonstrates a concentrated distribution toward lower tracking speed values, with a smooth spread across mid-range values—reflecting a natural and controlled movement experience.}
    \Description{This figure shows histograms of tracking speed usage across all tasks for Go-Go, PRISM, and ForcePinch. ForcePinch demonstrates a concentrated and smoother spread, indicating fine-grained user control.}
    \label{fig:tracking-hist}
\end{figure}

\topic{Force input enables more responsive speed adjustments.}
As shown in Figure \ref{fig:tracking-hist}, we compared the tracking speed histograms of ForcePinch, Go-Go, and PRISM across all three tasks (Constant was excluded due to its fixed tracking speed). In both Task 1 and Task 2, ForcePinch allowed users to access the lowest tracking speed ranges more frequently, supporting higher precision during interaction. Participants echoed this in interviews, noting that ``when I want to reduce the tracking speed, I simply adjust the pinching force. This process feels more direct and responsive than modifying distance (Go-Go) or adjusting movement speed (PRISM).''

\begin{figure}
    \centering
    \includegraphics[width=\linewidth]{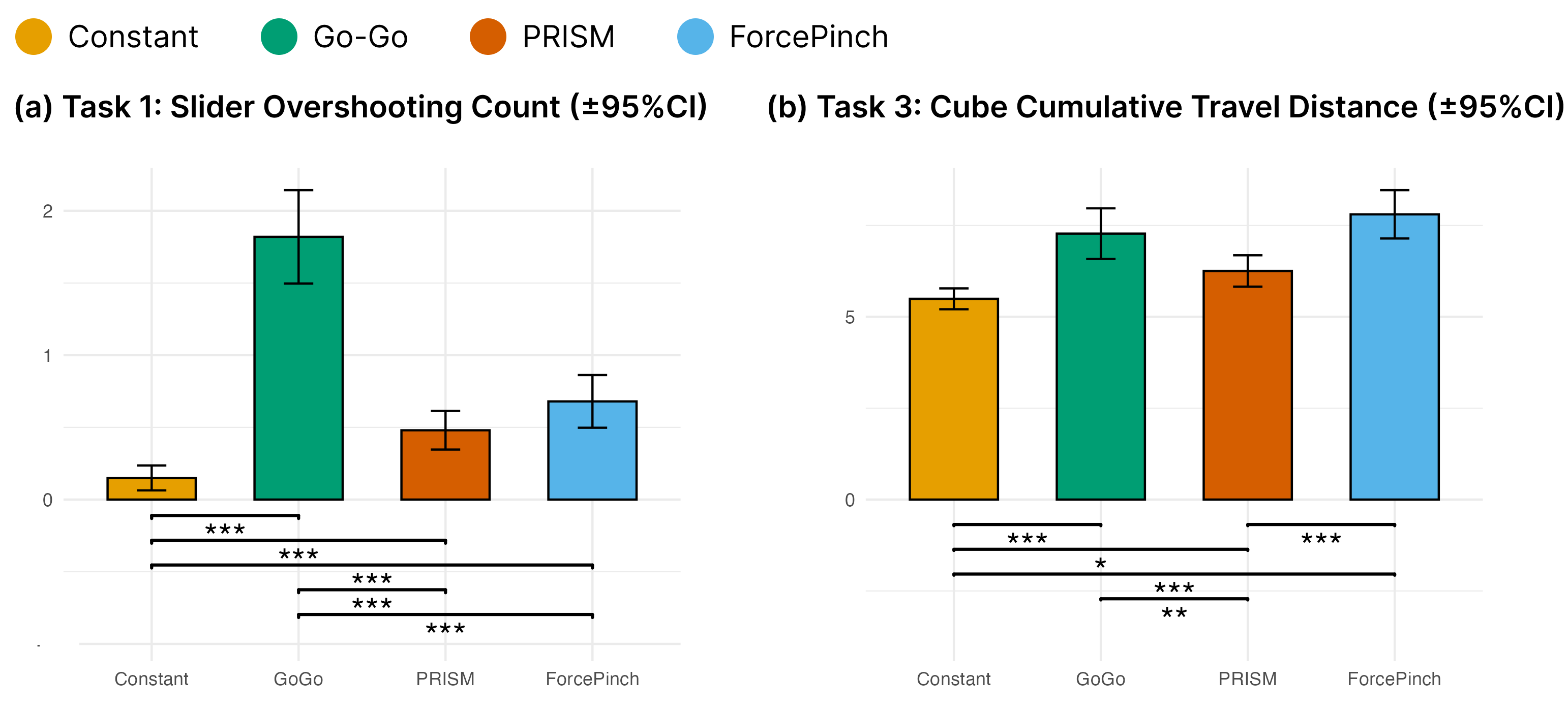}
    \caption{Overshooting in Task 1 and Task 3. We measured the number of overshooting instances in Task 1 (defined as moving beyond the target by more than 10\% of the intended distance) and the total distance traveled by the manipulated object (cube) in Task 3. Results show that ForcePinch led to significantly more overshooting or movement compared to the Constant method in both tasks (both ***, $p<0.001$).}
    \Description{This figure visualizes overshooting metrics for Tasks 1 and 3. ForcePinch leads to more overshooting or object travel distance than the Constant method, especially during rapid movement phases.}
    \label{fig:overshooting}
\end{figure}

\topic{Force is better suited for deceleration than acceleration.}
In both Task 1 and Task 3, we observed more frequent occurrences of the overshooting phenomenon than Constant when using ForcePinch, as illustrated in Figure \ref{fig:overshooting}. This suggests that users had less control during acceleration phases—when lighter force corresponds to higher tracking speeds. Interestingly, this stands in contrast to the precision benefits that ForcePinch offers. A possible explanation lies in the amplified movement required during acceleration: to move quickly, users must maintain fine control over larger hand motions, which imposes a higher cognitive load. This increased load may detract from users' attention to force control, making fast movements harder to regulate.
In contrast, during deceleration (i.e., heavier force, lower tracking speed), hand movements are scaled down, offering a greater margin for error. This allows users to focus more effectively on both tracking speed and hand movement, facilitating more precise control. This contrast suggests that ForcePinch is particularly well-suited for tasks requiring deceleration and fine-grained adjustments, rather than rapid, large-scale movements.

\topic{The impact of decoupling tracking speed from hand movement.} 
Unlike Go-Go and PRISM, ForcePinch decouples tracking speed from hand movement (either distance or velocity), introducing an additional degree of input freedom. While this added control enables more precise control and flexible interaction, it also increases cognitive load by introducing a separate control dimension. To address this, supportive design elements such as real-time visual feedback may be necessary to help users understand and manage their input, ultimately reducing cognitive demand.

\subsection{Why Reduced Effectiveness in Spatial 3D Placement Tasks?}

\topic{Two-Phase Strategy Leading to Spatial Overshoot.}
Participants frequently adopted a two-phase manipulation strategy in 3D tasks: a coarse, initial movement to reach the general target area, followed by finer movements for precise alignment. With ForcePinch, the initial coarse movement often resulted in overshooting, necessitating additional adjustments or regrasping. Overshooting is particularly problematic in 3D environments because open spatial contexts do not constrain object movement, potentially causing objects to move far away and making subsequent adjustments more difficult.

\topic{Limited Precision Without Numerical Feedback.}
ForcePinch showed no distinct precision advantage over other interaction methods in 3D tasks, largely due to the absence of numerical or explicit positional feedback. Unlike simpler 1D tasks, assessing precision in 3D placement requires participants to observe the object from multiple viewpoints. Reliance solely on visual feedback thus made error detection and correction less intuitive and more complex.

\topic{Mismatch Between Force Application and Spatial Movement Dynamics.}
Natural 3D movements typically exhibit an ``ease-in, ease-out'' dynamic \cite{flash1985coordination, wong2021energetic}, where users instinctively accelerate and apply more force to secure control over objects during rapid motion. However, ForcePinch inversely maps increased force to slower tracking speed, causing a mismatch with user expectations. This counterintuitive relationship, where lighter force corresponds to faster movements, likely contributed to the reduced effectiveness observed in 3D manipulation tasks.

\subsection{Establishing Expectations for Movement}

\topic{Establishing movement expectations is more difficult in 2D than in 1D.}
Due to the increased degrees of freedom and the continuous path-tracing requirement in Task 2, users needed more stable control compared to Task 1 (1D) and Task 3 (3D). However, perceiving hand movement—particularly in terms of distance and velocity—is more cognitively demanding in 2D. This challenge was especially evident with the Go-Go technique, where estimating the distance between the cursor and its starting point on the canvas proved unintuitive. As a result, users struggled to anticipate changes in tracking speed, making it difficult to maintain stable interaction throughout the task.

\begin{figure}
    \centering
    \includegraphics[width=\linewidth]{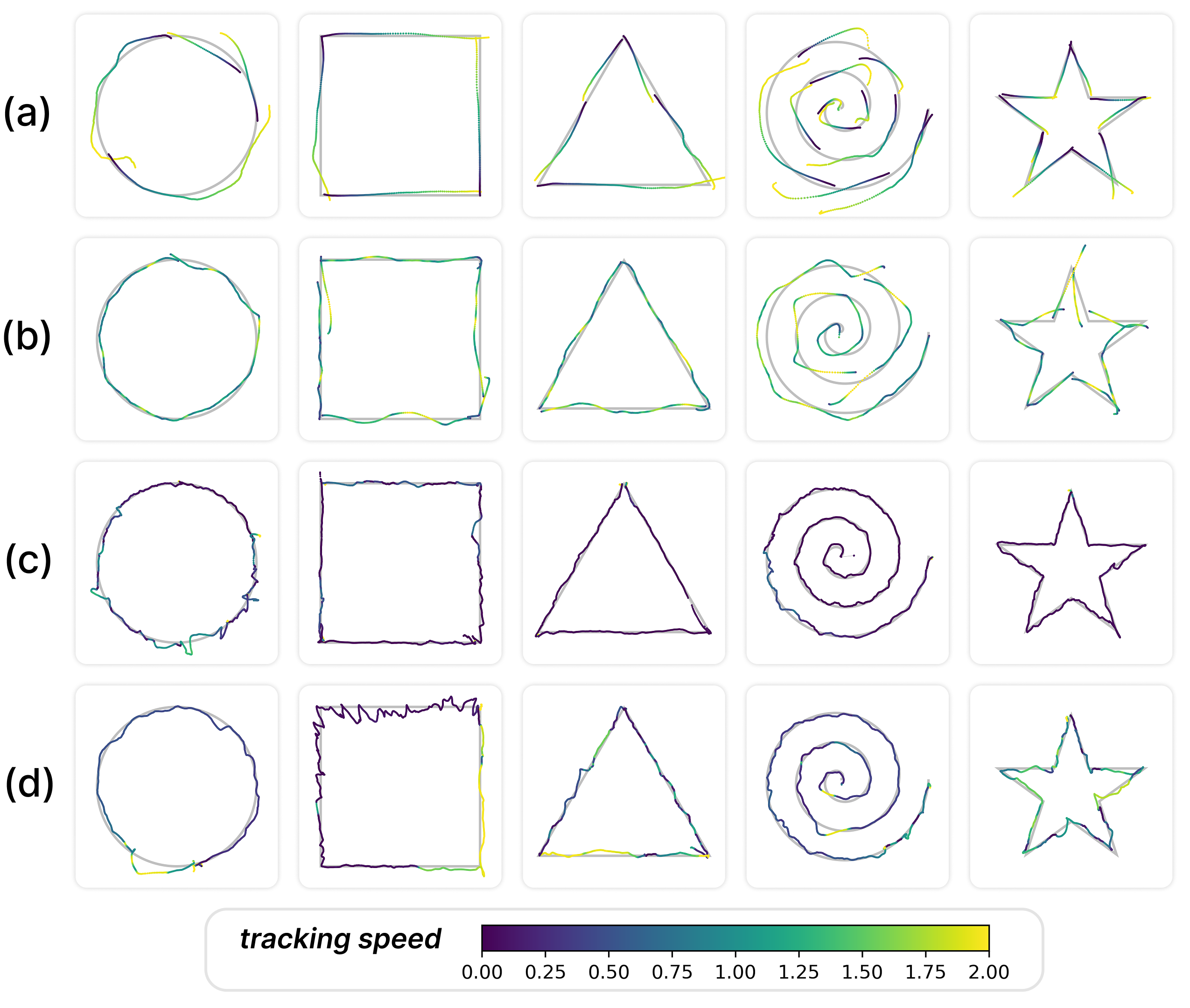}
    \caption{Representative Drawings displaying tracking Speed variation in Task 2: (a) Go-Go, (b) PRISM, and (c, d) ForcePinch.}
    \Description{This figure presents example drawings in the 2D tracing task. (a) and (b) show variable speed performance of Go-Go and PRISM; (c) and (d) show two ForcePinch usage strategies—constant low speed vs. dynamic modulation.}
    \label{fig:task2-tracking}
\end{figure}

This lack of predictability led to noticeable instability in Go-Go's performance in Task 2 (Figure \ref{fig:task2-tracking}a), with users frequently over- or under-shooting their paths, especially in the spiral and star shapes, where there are many changes in direction. Similarly, PRISM showed fluctuating tracking speeds throughout the interaction (Figure \ref{fig:task2-tracking}b), even in the square and triangle paths, geometries that typically favor consistent movement. In contrast, ForcePinch decouples tracking speed from physical hand movement, shifting the control mapping from ``hand movement $\rightarrow$ tracking speed'' to a simpler and more direct ``force $\rightarrow$ tracking speed.'' This reduction in cognitive complexity allowed users to form clearer expectations, resulting in smoother and more stable pointer control.

\topic{ForcePinch enables more flexible expectation strategies.}
During Task 2, we observed two distinct usage patterns among participants using ForcePinch. Some users chose to keep their fingers tightly pinched throughout the task, maintaining a consistently low tracking speed. In doing so, they effectively used ForcePinch as a slower but more precise variant of the Constant method, as illustrated by the dark lines in Figure \ref{fig:task2-tracking}c. This was especially effective for curved shapes like the circle and spiral, where precision was prioritized over speed.
Other users adapted their force dynamically: applying lighter pressure when drawing straight lines and switching to stronger force when navigating curves or corners. This allowed them to modulate speed and precision in response to the demands of different path segments (Figure \ref{fig:task2-tracking}d). This is most noticeable in shapes like the star and triangle, which contain many corners. These distinct strategies highlight ForcePinch's flexibility, offering users the freedom to choose an interaction style that best matches their goals and preferences.

\topic{Movement expectations are easier to form in 3D scenarios.}
Interestingly, Go-Go and PRISM both showed improved performance and user experience in the 3D task (Task 3). The spatial nature of 3D environments may reduce cognitive load by making it easier for users to perceive and predict hand movement in space. This natural spatial awareness appears to support more intuitive control of tracking speed through physical motion. As a result, users could more easily form accurate expectations about how their movement would influence interaction outcomes.

\revision{

\begin{figure*}[htbp]
  \centering
  \begin{subfigure}[t]{0.22\textwidth}
    \includegraphics[width=\linewidth]{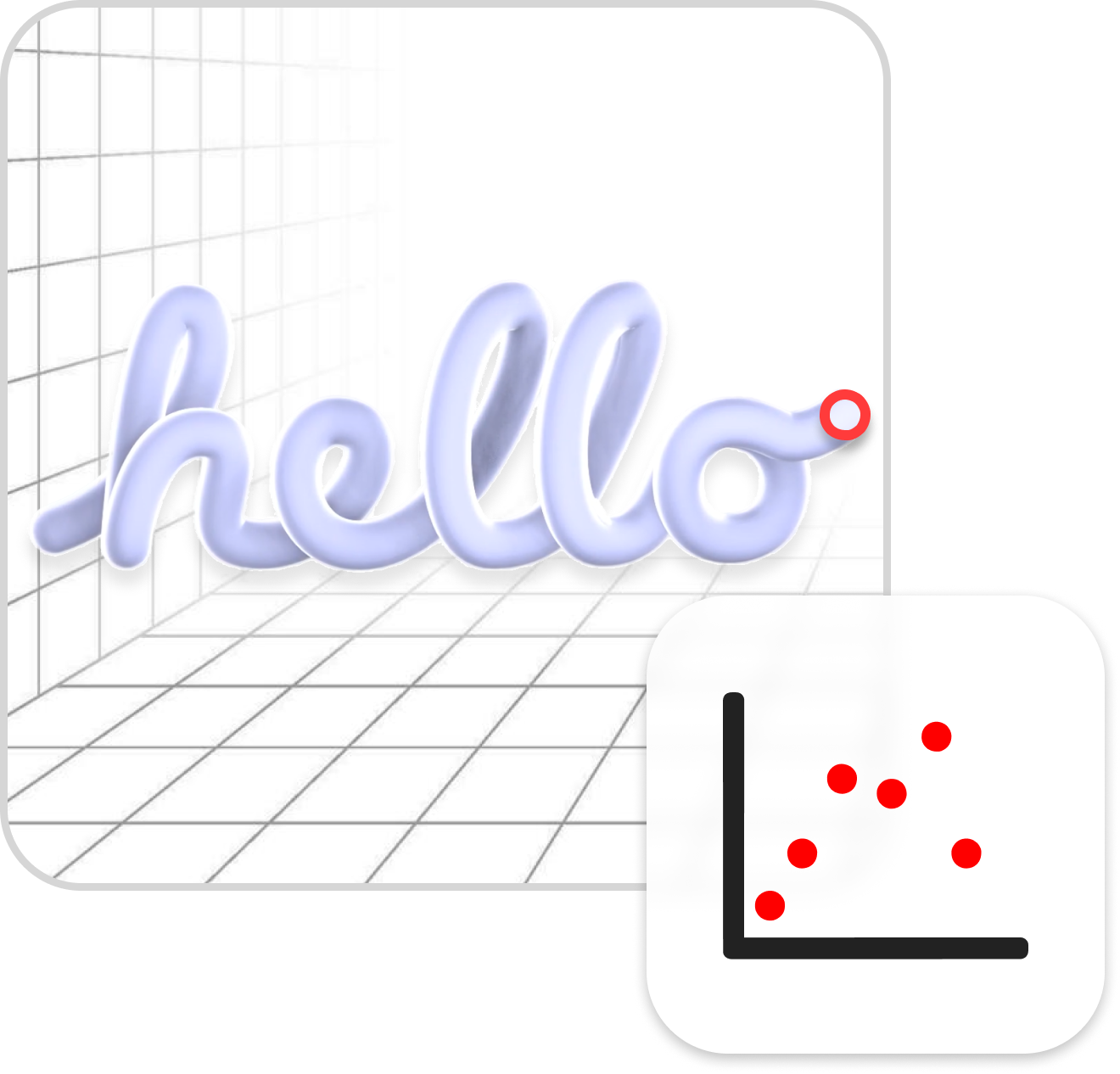}
    \caption{Magnitude (Absolute Value): \\3D Drawing.}
    \label{fig:magnitude}
  \end{subfigure}
  \hfill
  \begin{subfigure}[t]{0.22\textwidth}
    \includegraphics[width=\linewidth]{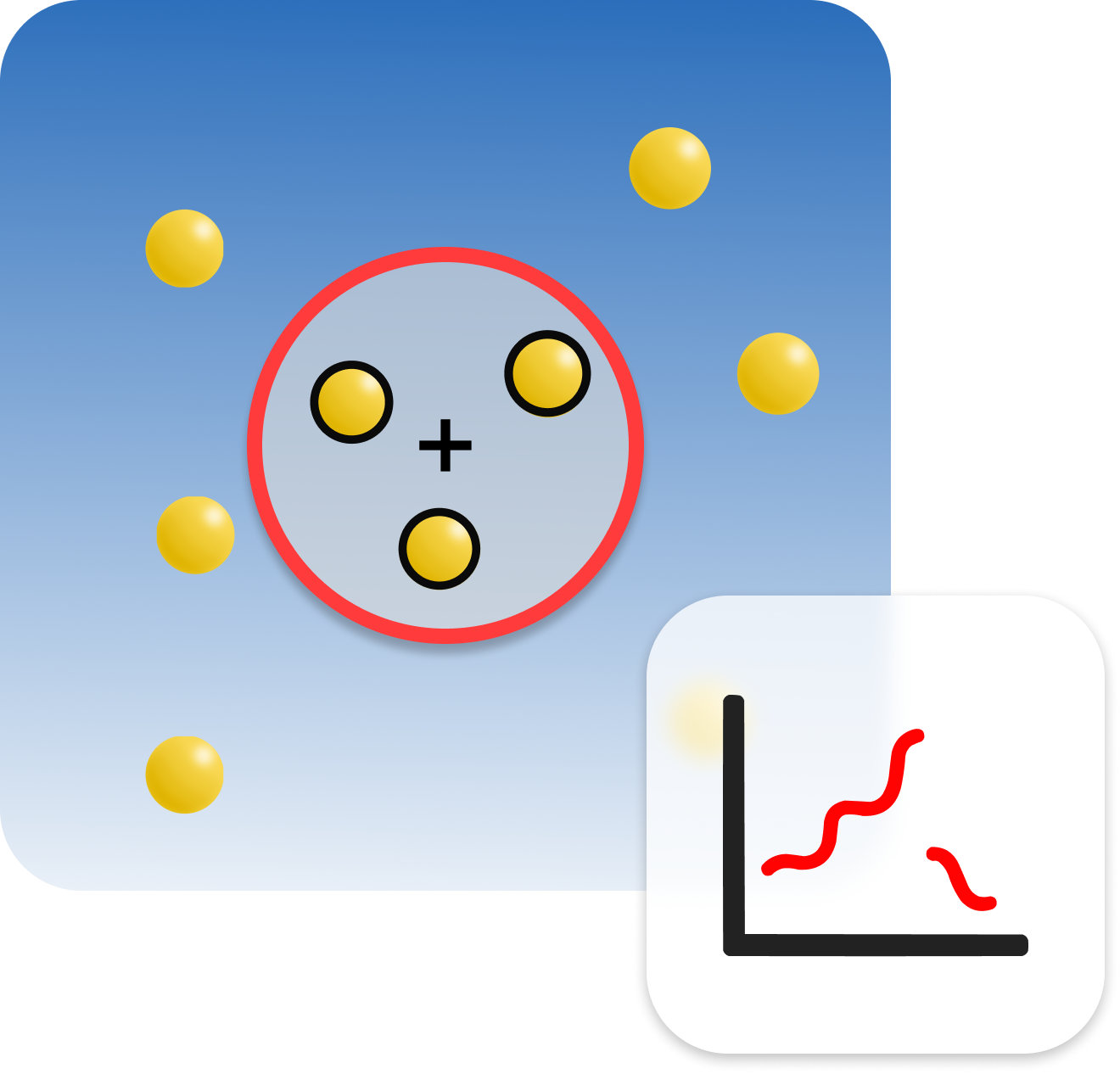}
    \caption{Change (Relative Value): \\Bubble Selection.}
    \label{fig:change}
  \end{subfigure}
  \hfill
  \begin{subfigure}[t]{0.22\textwidth}
    \includegraphics[width=\linewidth]{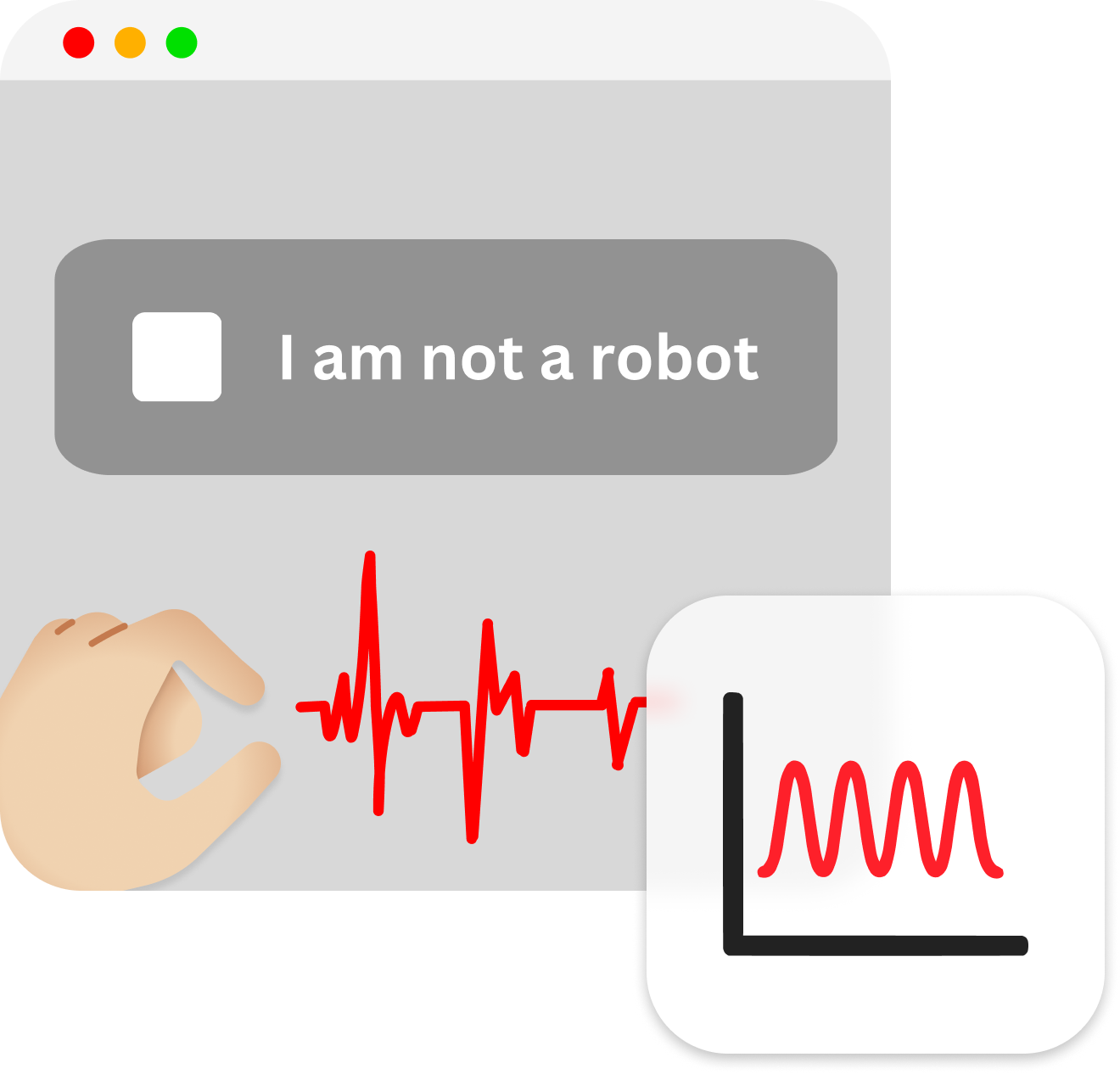}
    \caption{Rhythm (Temporal Pattern): \\Captcha and Signature.}
    \label{fig:rhythm}
  \end{subfigure}
  \hfill
  \begin{subfigure}[t]{0.22\textwidth}
    \includegraphics[width=\linewidth]{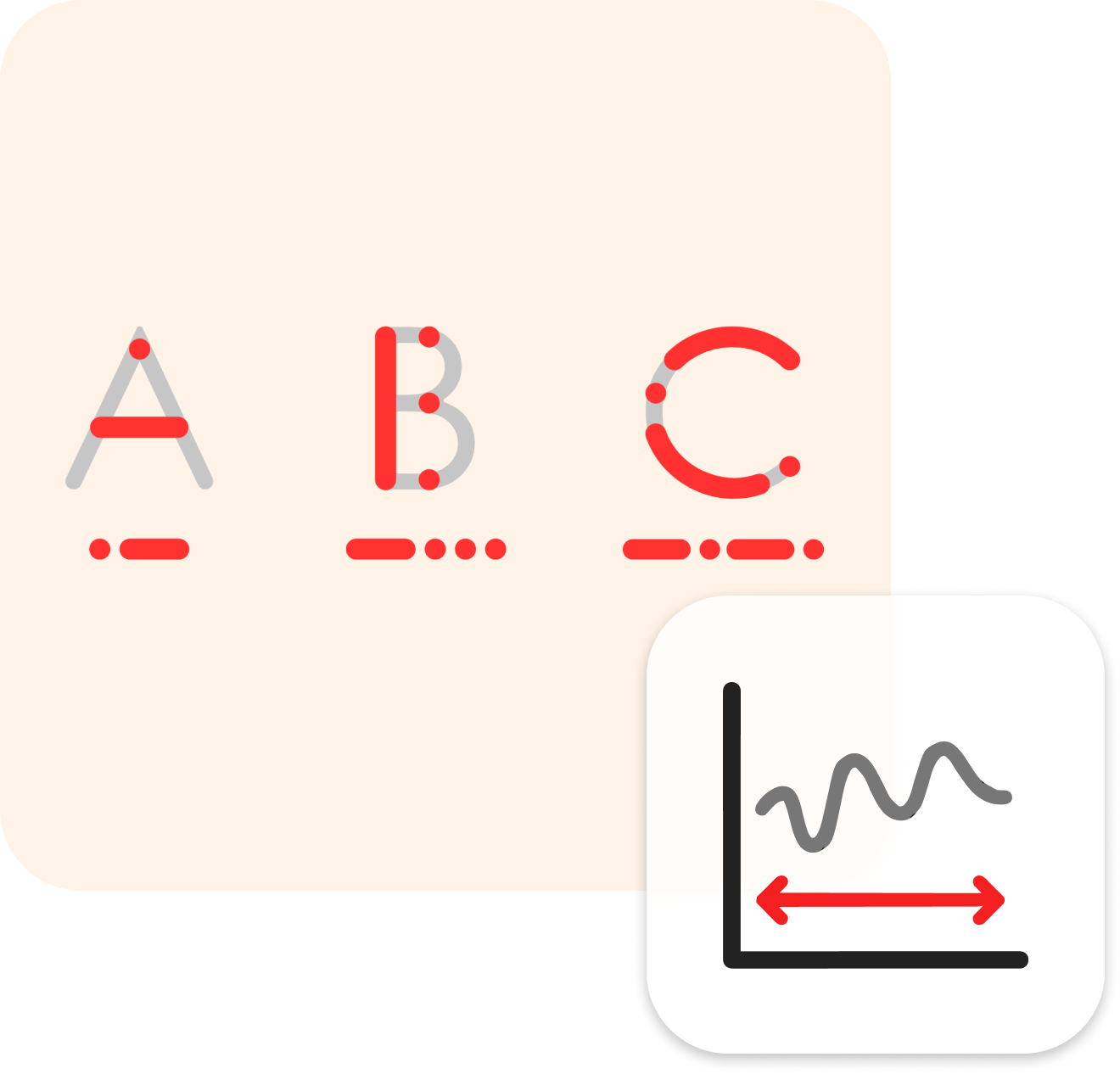}
    \caption{Duration (Timing): \\Morse Code Input.}
    \label{fig:duration}
  \end{subfigure}
  \vspace{1.5em}
  \caption{\revision{Four examples illustrating distinct interpretations of continuous force input in spatial interaction, each mapping a unique force signal characteristic.}}
  \Description{This figure illustrates four conceptual mappings of continuous force input: (a) Magnitude (3D drawing), (b) Change (bubble selection), (c) Rhythm (captcha/signature input), and (d) Duration (Morse code entry).}
  \label{fig:force-examples}
\end{figure*}

\section{Reflections and Future Work} \label{sec:reflection}

\subsection{Limitations}

\topic{Need for better calibration methods.}
While the current calibration process enabled all participants to complete the tasks, observations revealed notable limitations. Participants often applied excessive force during calibration to ensure system recognition, which may contribute to fatigue over time. Some struggled to differentiate force levels, leading to inaccurate mappings that compressed the effective input range and degraded system responsiveness. Since the force-to-speed mapping directly impacts usability, improving calibration is crucial. Future work could explore machine learning models to better capture individual force-response curves, leverage perceptual and physiological insights, and introduce real-time feedback during calibration. Hardware improvements, such as using more sensitive sensors like EMG, may also enhance stability and comfort.

\topic{Initial errors introduced during interaction triggering.}
We observed that interactions often began with minimal force—mapped to maximum tracking speed—which made small hand tremors more pronounced and led to initial jitter. This was especially detrimental in precision tasks like 2D tracing, where small early errors accumulated and increased task completion time. To address this, future systems could introduce a trigger threshold, requiring a minimum force to activate interaction. This would help avoid unintentional movements caused by low-force instability and improve control during the critical onset of interaction.

\subsection{Contextual Meaning of Force}

Our user study with ForcePinch revealed that users' interpretations of force behavior can vary significantly depending on context. For example, the ``friction'' metaphor was widely understood and intuitive for 1D or 2D cursor movement, where more force led to slower movement. In contrast, for 3D object manipulation (e.g., grabbing and moving), metaphors like ``centrifugal force'' felt more appropriate and better matched some participants' mental models. These observations underscore the importance of context when designing force-responsive interactions and suggest that adaptable, context-aware mappings can improve usability.

Moreover, user expectations regarding the effect of increased force were not uniform. Some users assumed that applying more force should accelerate movement or action (direct mapping), while others expected it to slow down interaction (inverse mapping). This divergence suggests that there is no universally intuitive mapping and highlights the importance of offering customization options, enabling users to select mappings that align with their personal preferences and cognitive models.

\subsection{Force-Responsive Interaction}

Pinch gestures are widely used for selection and confirmation in XR, and recent research has moved beyond binary detection to explore more nuanced pinch states, such as half-pinch~\cite{khushman2025hhe,kim2025pinchcatcher}. Building on this trend, our work leverages continuous pinch force as an expressive input channel. Our evaluation confirms that users can naturally modulate this force, enabling richer and more controllable interaction techniques.

While our implementation focuses on tracking speed control, we believe the underlying idea of continuous force input can be generalized to a broader set of interactions. To support and encourage this line of exploration, we introduce the concept of \textbf{force-responsive interaction}: embodied interactions that integrate continuous force modulation as a secondary input to augment or control a primary task. Force is a proprioceptively accessible, analog property that users can intuitively vary and perceive. When combined with spatial gestures, it enables more precise, expressive, and adaptable interactions across diverse XR contexts.

Conceptually, force-responsive interaction combines two interaction dimensions:

\begin{itemize}[leftmargin=*]
    \item \textbf{Primary Interaction}: The core embodied interaction task (e.g., selection, manipulation~\cite{zhang2024focusflow} and/or rotation~\cite{yang2020tilt}, creation~\cite{hertel2021taxonomy}).
    \item \textbf{Secondary Interaction (Force Mapping)}: How continuous force input is interpreted to influence the primary task.
\end{itemize}

For example, in \textit{ForcePinch}, users modulate tracking speed (C-D gain) while dragging objects by varying the magnitude of force.

Building on this idea, we identify four distinct ways in which continuous force input can be interpreted, illustrated in Figure \ref{fig:force-examples}. Each reflects a different characteristic of the time-varying force signal, revealing how a single modality—force—can support diverse and expressive interactions when thoughtfully mapped to tasks:

\subsubsection{\textbf{Magnitude (Absolute Value)}}
The raw intensity of force can directly modulate continuous parameters such as speed, size, or opacity.  
\textit{Example – 3D Drawing (Figure \ref{fig:change})}: Users draw in 3D by pinching with varying force to control stroke thickness—light force yields fine lines, strong force creates bold strokes. This enables fine-grained creative expression in spatial content creation~\cite{open-brush-docs}.

\subsubsection{\textbf{Change (Relative Value)}}
Changes in force—such as increasing, decreasing, or crossing thresholds—can trigger discrete events or mode shifts.  
\textit{Example – Bubble Selection (Figure \ref{fig:change})}: Applying more force expands a selection sphere to include nearby objects, while reducing force shrinks it. This allows dynamic selection without additional gestures, enhancing the primary selection task.

\subsubsection{\textbf{Rhythm (Temporal Pattern)}}
Rhythmic variations in force, such as tapping or pulsing patterns, support expressive, pattern-based inputs.  
\textit{Example – Captcha and Signature (Figure \ref{fig:rhythm})}: Unique force rhythms act as biometric patterns for identity verification or secure input. The inherent variability of human rhythm makes this approach robust against impersonation while remaining unobtrusive.

\subsubsection{\textbf{Duration (Timing)}}
Sustained force over time can differentiate actions or encode data.  
\textit{Example – Morse Code Input (Figure \ref{fig:duration})}: Users press and hold with different durations to input dots and dashes in Morse Code, enabling text entry without lifting their fingers. Duration-based input is intuitive and efficient for eyes-free or discrete input.

\subsection{Beyond Force\textit{Pinch}}

Although our study centers on force-enhanced pinch gestures, the underlying concept of force-responsive interaction is broadly applicable and not restricted to a specific input modality. Any body-based action that allows for variable force—such as clenching a fist, flexing a wrist, nodding, or microgestures~\cite{meta_microgestures}—can be used to instantiate this interaction model. The principle lies in augmenting primary input with a secondary force dimension, enabling richer control.

This idea extends naturally to multi-point or bimanual input. For instance, users could modulate different parameters—such as scale and rotation—using coordinated force from both hands or feet. Such configurations increase the expressive bandwidth, allowing for more nuanced and simultaneous control over multiple dimensions.

Our framework also supports diverse sensing technologies. In addition to finger-mounted sensors, force input can be detected using electromyography (EMG) signals~\cite{zhang2024may}, vision-based estimation methods~\cite{mollyn2024egotouch, grady2024pressurevision++}, smart textiles~\cite{luo2021learning}, or embedded sensors in wearables like smartwatches~\cite{buddhika2019fsense}. As these technologies mature—particularly those that are unobtrusive or passive—the feasibility and accessibility of force-responsive interaction will improve.

Finally, this approach can be readily integrated into existing commercial hardware. For example, enhancing the buttons or triggers on VR controllers with force-sensing capabilities would allow for more expressive operations, such as pressure-modulated selection or force-sensitive object manipulation in 3D space.

}
\section{Conclusion}

In this paper, we introduced \textbf{ForcePinch}, a force-responsive tracking speed control method that enables users to smoothly transition between rapid and precise movements by varying their pinching force. Through a user study with 20 participants across 1D, 2D, and 3D tasks, we found that ForcePinch supports more precise manipulation and offers a better user experience, particularly in tasks requiring fine-grained control.
Compared to prior techniques such as the distance-based Go-Go and velocity-based PRISM methods, ForcePinch provides greater flexibility by decoupling tracking speed from hand movement. Our findings also reveal interesting dynamics in how users interpret and apply force—such as diverging expectations about the effects of increased pressure—highlighting the nuanced role of force as an input dimension.
\revision{While our implementation centers on tracking speed, we reflect on how continuous force input may be generalized to enrich other embodied interactions. We describe this broader perspective as \textbf{force-responsive interaction}, where force acts as a secondary input channel to augment tasks like selection, manipulation, and creation. We outline key force signal characteristics—magnitude, change, rhythm, and duration—and show how each can support expressive interaction techniques. We hope that ForcePinch and this conceptual framing will inspire further exploration of force-enhanced interaction in mixed reality.}

\begin{acks}
We thank Wanbo Geng for illustrating the figures in the reflection section. This work was supported in part by NSF grant IIS-2441310 and the Google Research Scholar program.
\end{acks}

\bibliographystyle{ACM-Reference-Format}
\bibliography{cite}

\appendix
\begin{figure}
    \centering
    \begin{subfigure}[b]{1.0\linewidth}
        \centering
        \includegraphics[width=\linewidth]{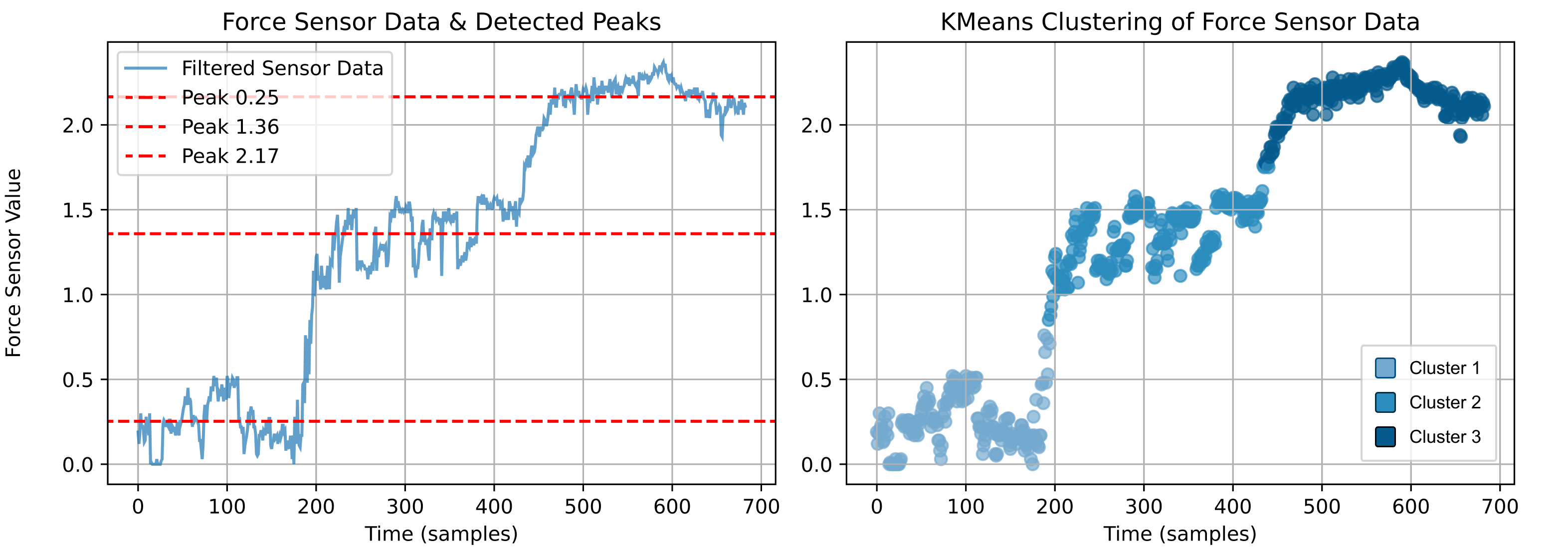}
        \caption{Force Sensor Data Processing: Detection of characteristic force levels (left) through k-means clustering (right).}
        \label{fig:kmeans}
    \end{subfigure}
    
    \vspace{1.2em} 

    \begin{subfigure}[b]{1.0\linewidth}
        \centering
        \includegraphics[width=\linewidth]{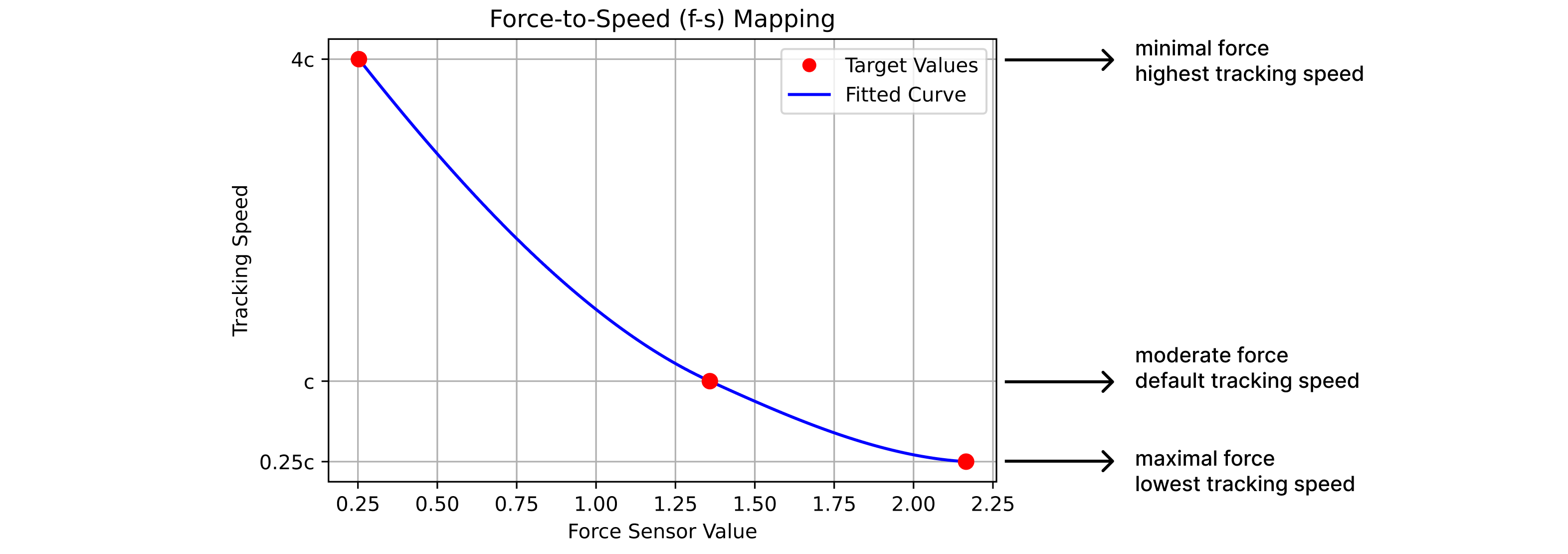}
        \caption{Force-Speed Mapping Curve: Calibration function using cubic Hermite spline interpolation to map raw force sensor values to a smooth tracking speed output.}
        \label{fig:f-s_mapping_data}
    \end{subfigure}
    \vspace{0.8em}
    \caption{Individualized Calibration. (a) captures user-applied force levels through clustering, identifying three characteristic force points. (b) constructs a smooth mapping curve to distribute force input across the tracking speed range. This process accounts for individual differences in strength and perception, and ensures consistent, user-specific experience.}
    \Description{This figure shows the individualized calibration process. (a) illustrates force level detection using k-means clustering. (b) plots the resulting force-to-speed curve using cubic Hermite spline interpolation to personalize tracking speed control.}
    \label{fig:calibration}
\end{figure}

\section{Individualized Calibration}  \label{sec:calibration}

Due to individual differences in both force application and perception, we perform a personalized calibration for each user. This calibration captures three key parameters: (1) the user-defined \emph{moderate force}, which serves as a neutral reference point for modulating tracking speed; (2) the \emph{maximum} and \emph{minimum perceivable forces}, which define when the slowest and fastest tracking speeds should be applied, respectively; and (3) the full range between these force extremes.

As illustrated in Figure~\ref{fig:kmeans}, the calibration procedure begins with the user applying three distinct levels of pinch force, ranging from their minimum to maximum comfortable values. Each level is held for approximately one second, during which we collect continuous time-series data of the applied force. We then apply k-means clustering ($k=3$) to identify three representative force levels: minimal, moderate, and maximal.

These three force levels are then mapped to the tracking speed control space: minimal force corresponds to the highest tracking speed, moderate force to the default (unmodified) tracking speed, and maximal force to the lowest tracking speed. These anchor points define two continuous intervals, over which we construct a smooth force-to-speed ($f-s$) mapping function. To ensure a more uniform and natural control experience, we use cubic Hermite spline interpolation instead of simple linear interpolation, as shown in Figure~\ref{fig:f-s_mapping_data}. For force values falling outside the calibrated range, tracking speed is clamped to the nearest boundary (i.e., maximum or minimum speed).

\end{document}